\documentclass[10pt,a4paper,notitlepage]{article}
\usepackage{setspace}
\usepackage{varioref}
\usepackage{multicol}
\usepackage[ansinew]{inputenc}
\usepackage{ulem}
\usepackage{paralist}
\usepackage{multirow}
\usepackage{array}
\usepackage{booktabs}
\usepackage{hyperref}
\usepackage[hang,footnotesize,bf]{caption}
\usepackage{graphicx}
\usepackage{epstopdf}
\usepackage{sidecap}
\usepackage{subfig}
\usepackage{wrapfig}
\DeclareGraphicsExtensions{.pdf,.eps,.ps,.png,.jpg}
\usepackage{pifont}
\usepackage{marvosym}
\usepackage{url}
\usepackage{slashed}
\usepackage[intlimits]{amsmath}
\usepackage{amsfonts}
\usepackage{amssymb}
\usepackage{amsthm}
\usepackage{mathrsfs}
\usepackage{moreverb}
\renewcommand{\listinglabel}[1]{\llap{\footnotesize\rmfamily\the#1:\s}}

\usepackage{algorithmic}
\usepackage[ruled,section]{algorithm}

\usepackage{boxedminipage}
\usepackage[version=3]{mhchem}

\newtheorem*{dfn*}{Definition}

\newcommand{\s}{\;\:}


\begin{document}
\begin{flushleft}
\vphantom{title}
\end{flushleft}
\begin{center}
{\Large Setting limits on supersymmetry using simplified models}\\[0.3cm]
C. G\"utschow$^1$\footnote{Available at \href{mailto:chris.g@cern.ch}{chris.g@cern.ch}}, Z. L. Marshall$^2$\footnote{Available at \href{mailto:zach.marshall@cern.ch}{zach.marshall@cern.ch}}\\[0.2cm]
$^1$University College London; $^2$CERN\\[0.2cm]
\end{center}
%
\rule{\linewidth}{0.1mm}
%
%
\begin{abstract}
Experimental limits on supersymmetry and similar theories are difficult to set because of the enormous available parameter space and difficult to generalize because of the complexity of single points.
Therefore, more phenomenological, simplified models are becoming popular for setting experimental limits, as they have clearer physical implications.
The use of these simplified model limits to set a real limit on a concrete theory has not, however, been demonstrated.
This paper recasts simplified model limits into limits on a specific and complete supersymmetry model, minimal supergravity.
Limits obtained under various physical assumptions are comparable to those produced by directed searches.
A prescription is provided for calculating conservative and aggressive limits on additional theories.
Using acceptance and efficiency tables along with the expected and observed numbers of events in various signal regions, LHC experimental results can be re-cast in this manner into almost any theoretical framework, including non-supersymmetric theories with supersymmetry-like signatures.
\end{abstract}
\rule{\linewidth}{0.1mm}
\begin{multicols}{2}
\thispagestyle{empty}
\renewcommand*\thesubsubsection{\Roman{subsubsection}}
\subsubsection{Introduction}
One of the most promising extensions of the Standard Model, supersymmetry (SUSY), is the central focus of many searches by the LHC experiments at CERN \cite{susy-1}--\cite{susy-7}. The data collected in 2011 is already sufficient to push the limits of new physics beyond those of any previous collider \cite{0lpaper}--\cite{cms4}. As new data arrive and the exclusions are pushed still farther, it will be increasingly important to clearly communicate to the physics community what regions of the extensive supersymmetric parameter space have been excluded. Current limits are typically set on constrained two-dimensional planes, which frequently do not represent the diverse available parameter space and are difficult to understand as limits on physical masses or branching fractions. A large set of simplified models~\cite{sidewalk,simplified-models} have been proposed for aiding in the understanding of these limits, and both ATLAS and CMS have provided exclusion results for several of these models~\cite{1lpaper,cms3,cms4}.  

This paper demonstrates the application of these simplified model exclusions to the minimal supergravity (mSUGRA, also known as the CMSSM) framework~\cite{msugra-1}--\cite{cmssm}.  This model is chosen in order to compare the limits set using simplified models to those published independently by the experiments.  In arriving at the exclusion curves, several tests are carried out to ensure that simplified models adequately describe the kinematics of complete mSUGRA parameter space points.  As this represents the first attempt to ``close the loop'' and set limits on SUSY using simplified models, a number of assumptions about the applicability of limits on particular simplified models are explored, resulting in recipes for setting conservative and aggressive limits on theories that have not been constrained by a direct search.

Section~\ref{sec:deconstruction} explains the deconstruction of a few of the complex mSUGRA points into their constituent pieces and discusses various trends across the mSUGRA phase space.  Section~\ref{sec:reconstruction} then discusses combining a few specific simplified models in order to re-create one of the complex points.  The kinematics are compared in order to evaluate the ability of the simplified models to reproduce the various features of a complex theory.  In Section~\ref{sec:limits}, several prescriptions for constructing a limit on a concrete theory are discussed, along with the physical and unphysical assumptions that lie behind each recipe.  Finally, in Section~\ref{sec:future}, other approximations are laid out in order to expand the applicability of the already-available simplified models to a larger range of theories.
\subsubsection{Model Deconstruction}
\label{sec:deconstruction}
Events are generated across several planes in this mSUGRA phase space in order to understand trends and evaluate the success of simplified models in covering the major features of the parameter space.  Mass spectra at a given point are generated using {\sc Isasugra}~\cite{isajet}.  The branching fractions are acquired using MSSMCalc, and matrix element event generation is done using {\sc MadGraph 5} 1.3.9~\cite{MadGraph} with CTEQ 6L1 PDFs~\cite{CTEQ}.  The events are then run through {\sc Pythia 6.425} for the decay of the supersymmetric particles, parton showering, and hadronization~\cite{Pythia}.  Because the LHC experiments use leading order generators for SUSY event production, inclusive matrix element matching is done (i.e. the parton shower is allowed to control all initial- and final-state radiation).  In order to mimic an LHC detector, the events are passed through the Pretty Good Simulation ({\sc PGS})~\cite{PGS} with an ``ATLAS'' parameter card, and cone jets with a radius of 0.4 are reconstructed using {\sc Fastjet}~\cite{fastjet}.  The agreement of this simulation with standard LHC simulations is discussed further in Section~\ref{sec:limits}.  NLO cross-sections for each point are calculated using {\sc Prospino} 2.1~\cite{Prospino}.  Each event is classified according to the topology of the matrix element and subsequent decay of the supersymmetric particles.  Light quarks, $u$, $d$, $s$, and $c$, as well as their supersymmetric partners, are all treated equally and independently from bottom and top quarks and sbottom and stop squarks.  

\begin{figure*}
\centering
\graphicspath{{plots/}}

\subfloat[$\tilde{q}$-dominated production mechanisms]{\label{fig:sq-prod}
\includegraphics[width=0.46\textwidth]{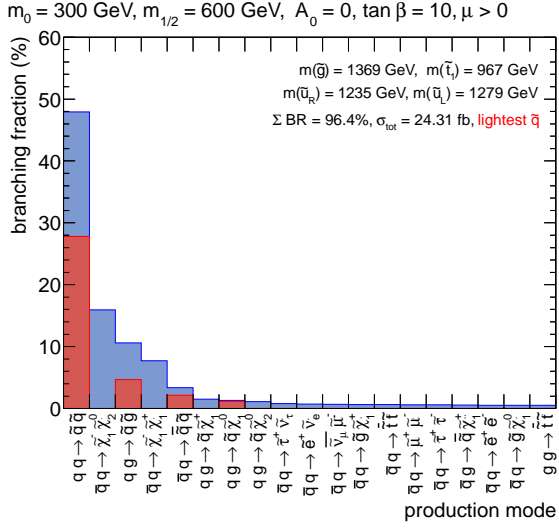}}
\quad\subfloat[$\tilde{q}$-dominated decay chains]{\label{fig:sq-dec}
\includegraphics[width=0.46\textwidth]{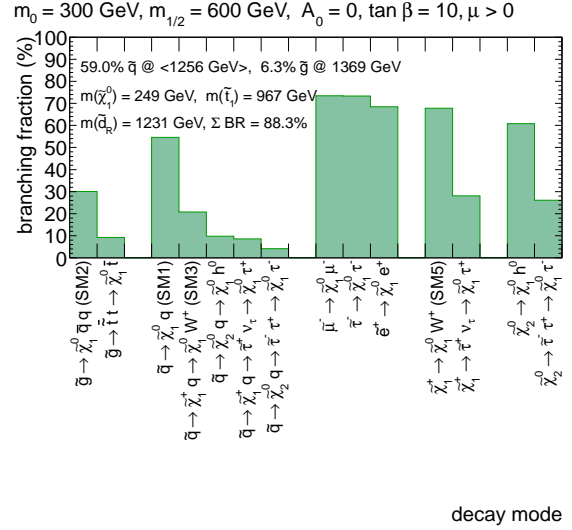}}
\linebreak\subfloat[production mechanisms in interjacent region]{\label{fig:both-prod}
\includegraphics[width=0.46\textwidth]{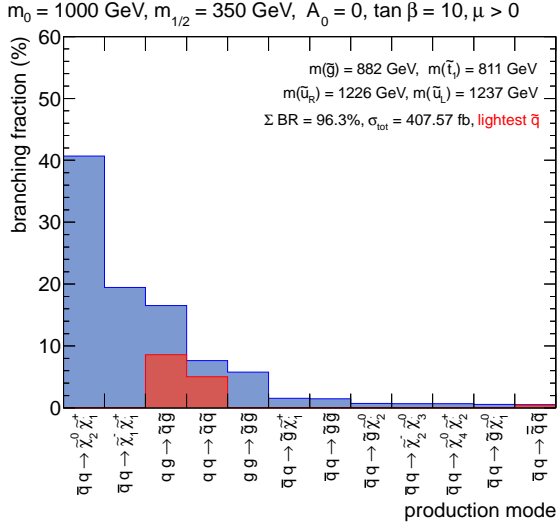}}
\quad\subfloat[decay chains in interjacent region]{\label{fig:both-dec}
\includegraphics[width=0.46\textwidth]{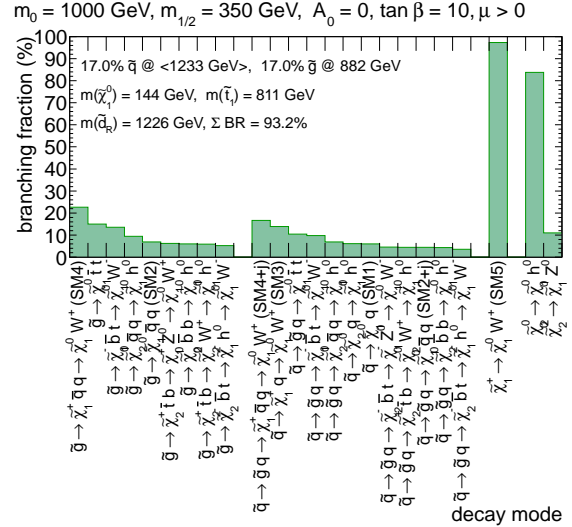}}
\caption{
Branching ratios for SUSY production mechanisms and decay modes in the mSUGRA parameter space. The top row ($m_0=300\,{\rm GeV}$, $m_{1/2}=600\,{\rm GeV}$, $\tan(\beta)=10$, $A_0=0$\,GeV, and $\mu>0$) is typical for the region in parameter space that is dominated by squark production, and the bottom row ($m_0=1000\,{\rm GeV}$, $m_{1/2}=350\,{\rm GeV}$, $\tan(\beta)=10$, $A_0=0$\,GeV, and $\mu>0$) is typical for the region in parameter space lying somewhat in between the two extremes. For clarity, production and decay modes are only listed if their branching fraction is greater than 0.5\,\%. The labels ``SM'' with a number are given to decay modes corresponding to the simplified models discussed in Section~\ref{sec:reconstruction}.
}
\label{fig:br-plots1}
\end{figure*}

\begin{figure*}
\centering
\graphicspath{{plots/}}
\subfloat[$\tilde{q}$-dominated production mechanisms]{
\includegraphics[width=0.46\textwidth]{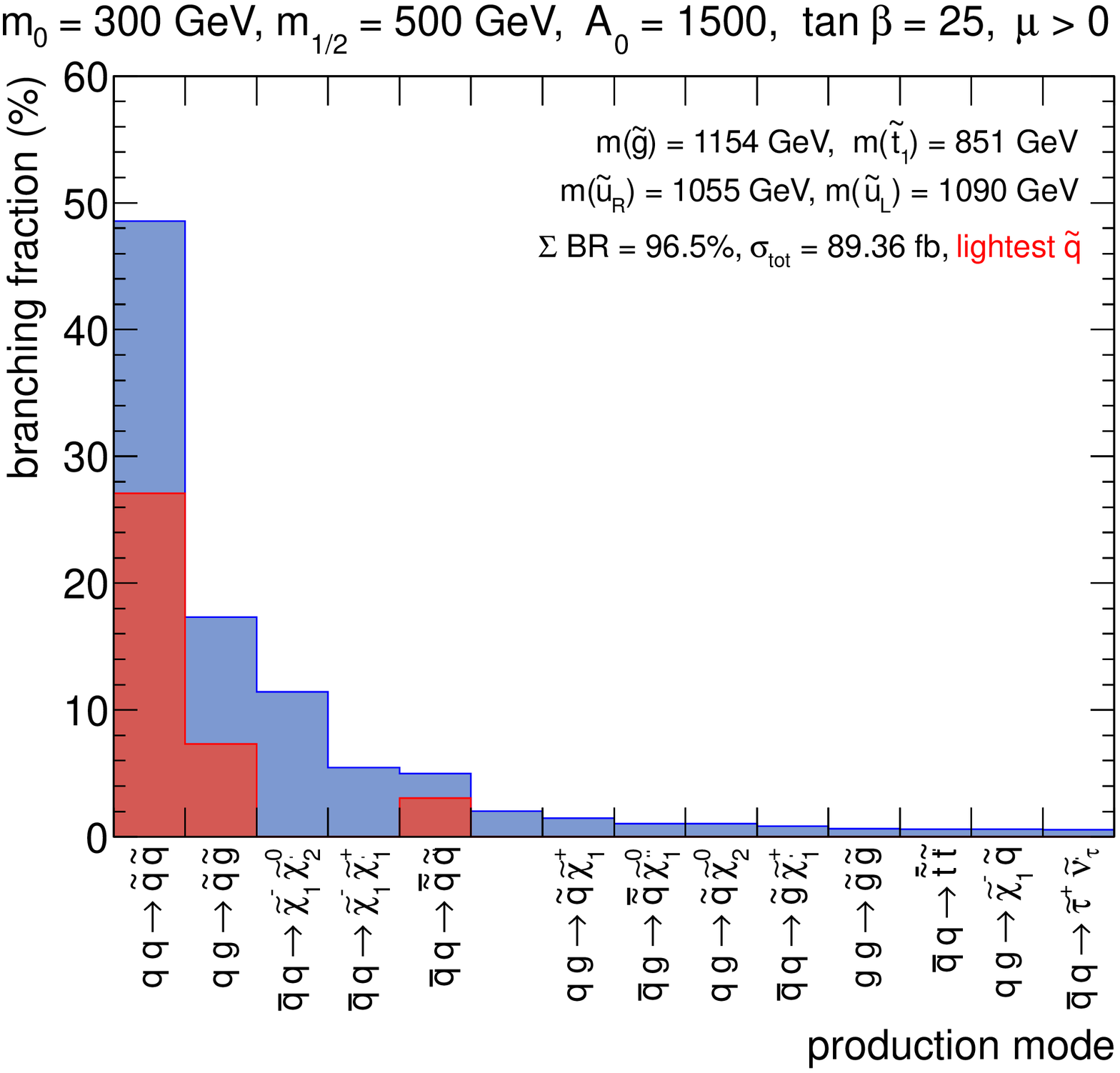}}
\quad\subfloat[$\tilde{q}$-dominated decay chains]{
\includegraphics[width=0.46\textwidth]{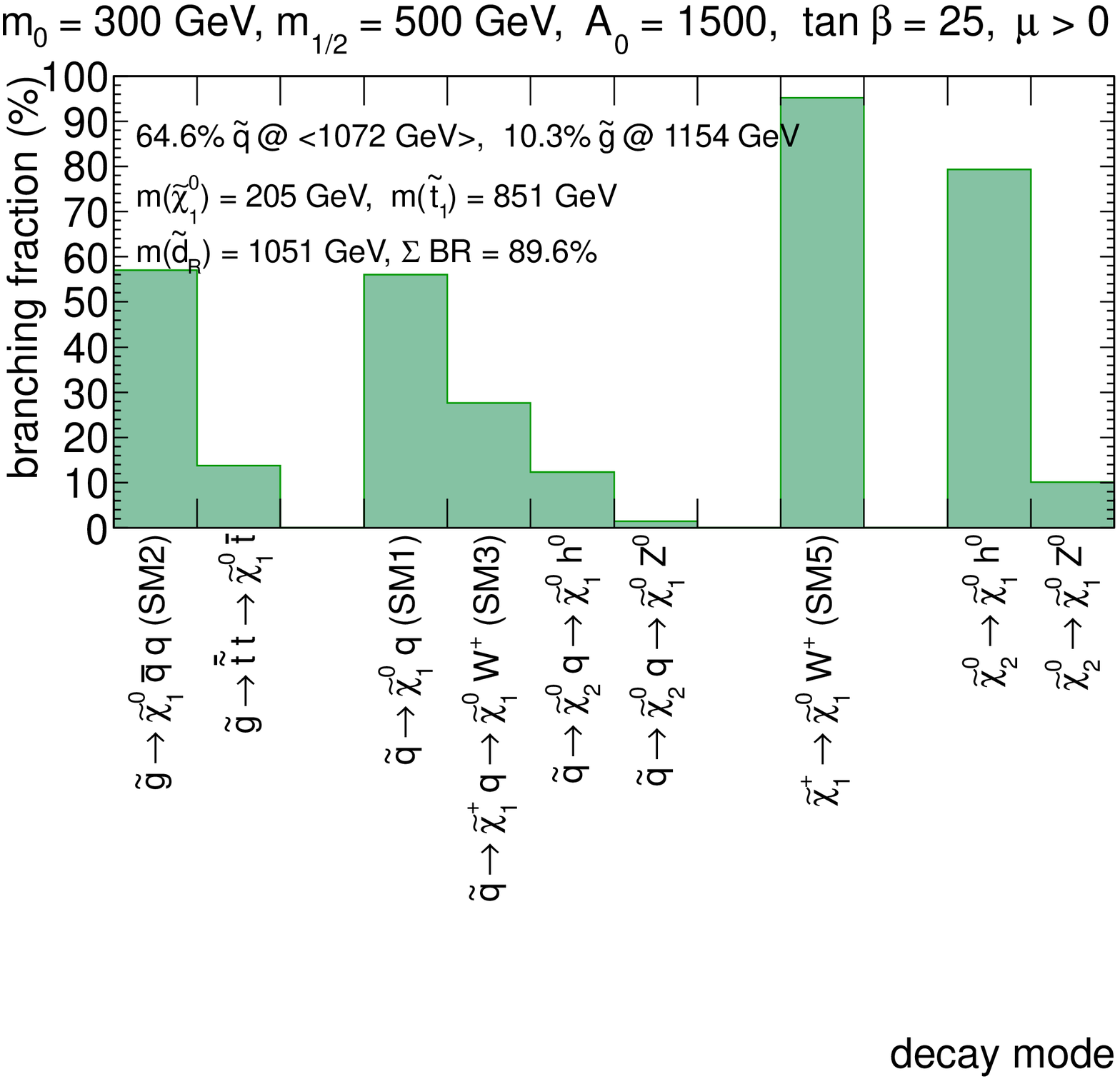}}
\linebreak\subfloat[$\tilde{g}$-dominated production mechanisms]{
\includegraphics[width=0.46\textwidth]{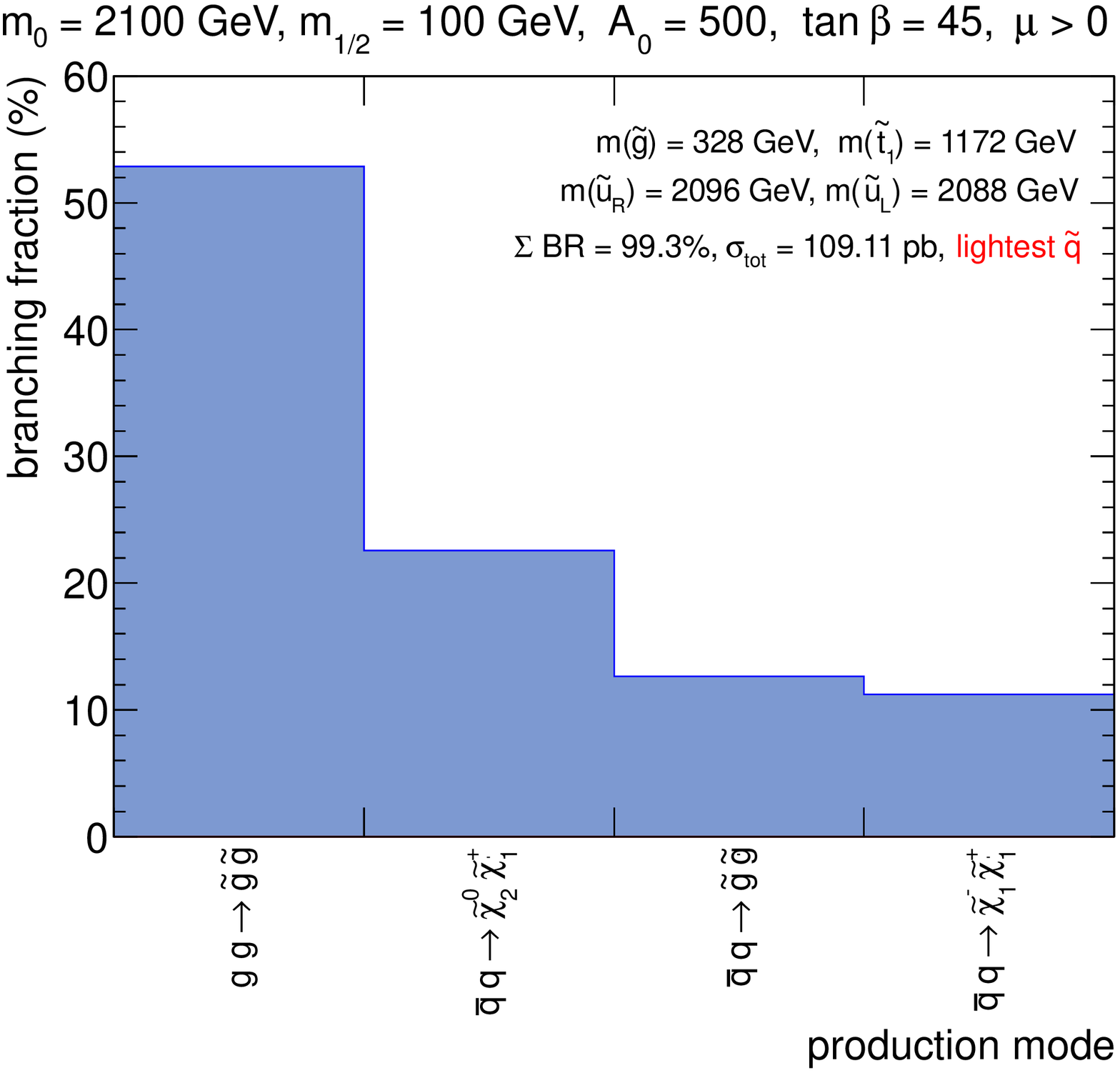}}
\quad\subfloat[$\tilde{g}$-dominated decay chains]{
\includegraphics[width=0.46\textwidth]{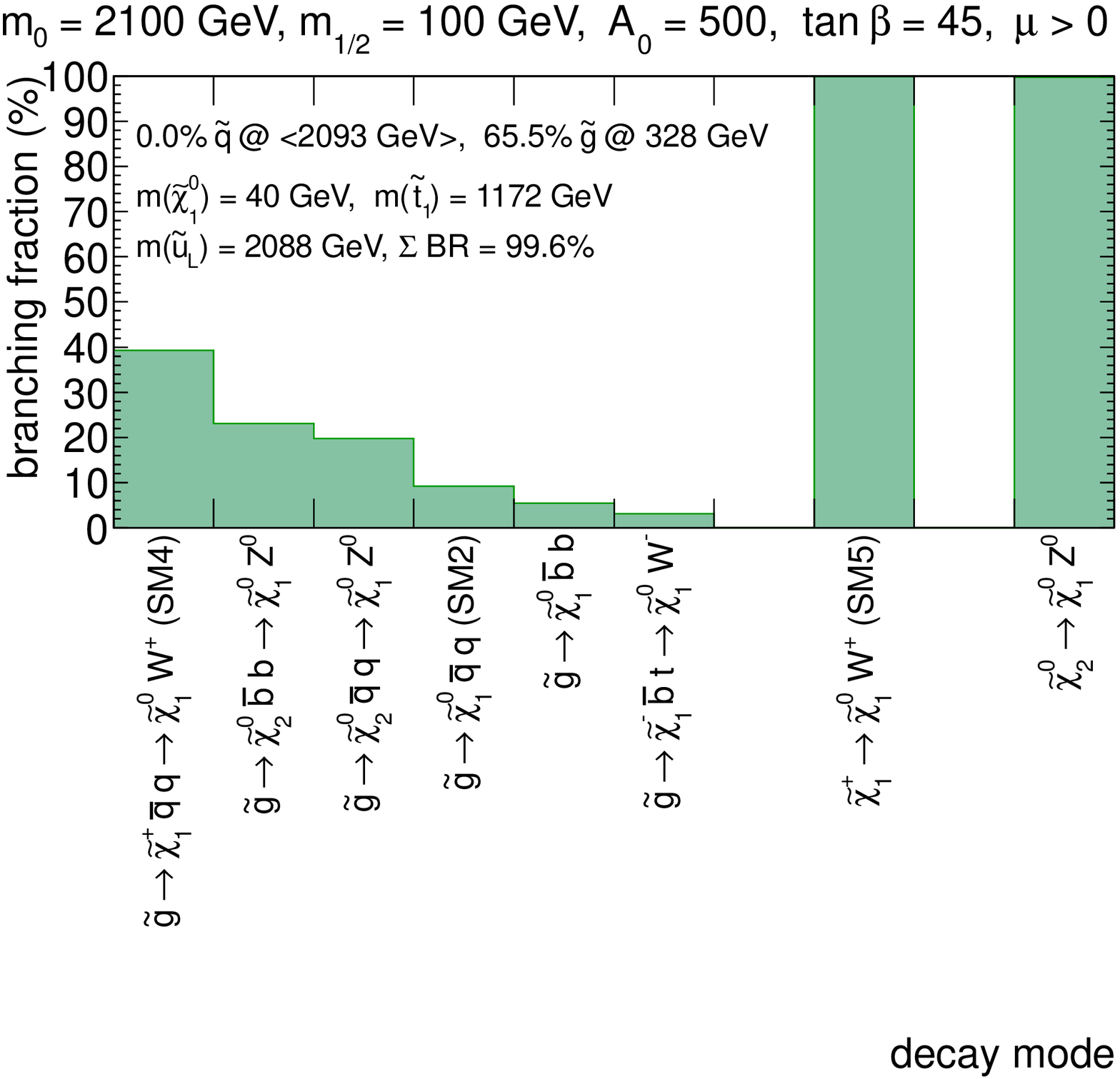}}
\caption{
Branching ratios for SUSY production mechanisms and decay modes in the mSUGRA parameter space. The top row ($m_0=300\,{\rm GeV}$, $m_{1/2}=500\,{\rm GeV}$, $\tan(\beta)=25$, $A_0=1500$ GeV, and $\mu>0$) is typical for the region in parameter space that is dominated by squark production, and the bottom row ($m_0=2100\,{\rm GeV}$, $m_{1/2}=100\,{\rm GeV}$, $\tan(\beta)=45$, $A_0=500$ GeV, and $\mu>0$) is typical for the region dominated by gluino production.  For clarity, production and decay modes are only listed if their branching fraction is greater than 0.5\,\%. The labels ``SM'' with a number are given to decay modes corresponding to the simplified models discussed in Section~\ref{sec:reconstruction}.
}
\label{fig:br-plots2}
\end{figure*}

The mSUGRA framework is characterized by four parameters and a sign: the unified gaugino and scalar particle masses $m_{1/2}$ and $m_0$, the unified trilinear coupling parameter $A_0$, the ratio of the vacuum expectation values of the two Higgs doublets $\tan(\beta)$ and the sign of the Higgsino mass parameter $\mu$.  Several examples of dominant production and decay modes for a few selected mSUGRA points are shown in Figures~\ref{fig:br-plots1} and \ref{fig:br-plots2}.  In these figures and elsewhere, only the immediate daughters of each supersymmetric particle are listed in the decay chain in order to shorten the representation of long decay chains.  Charge conjugate modes are combined.  Immediately, the complexity of typical supersymmetric points is visible.  Many production modes and decay modes may be open, producing a large number of possible event topologies.  This complexity makes limits on mSUGRA both difficult to understand in terms of masses and branching fractions and difficult to reinterpret and apply to alternate supersymmetric theories\footnote{
A large number of similar plots for other points in SUSY parameter space are available online at \href{http://cgutscho.web.cern.ch/cgutscho/susy/}{\texttt{cern.ch/christian/susy}}
}.  The classification of the SUSY decay modes in Figures~\ref{fig:br-plots1} and \ref{fig:br-plots2} into the simplified models discussed in Section~\ref{sec:reconstruction} is included in the legend.  When hard QCD radiation is produced in the event, the event is listed as a simplified model ``plus jets.''  Hard photon radiation is significantly less common and is included directly in the decay chain label.

\begin{figure*}
\centering
\graphicspath{{plots/}}
\subfloat[$\tilde{q}\tilde{q}$-production]{\label{fig:lo2d-sq-prod}
\includegraphics[width=0.47\textwidth]{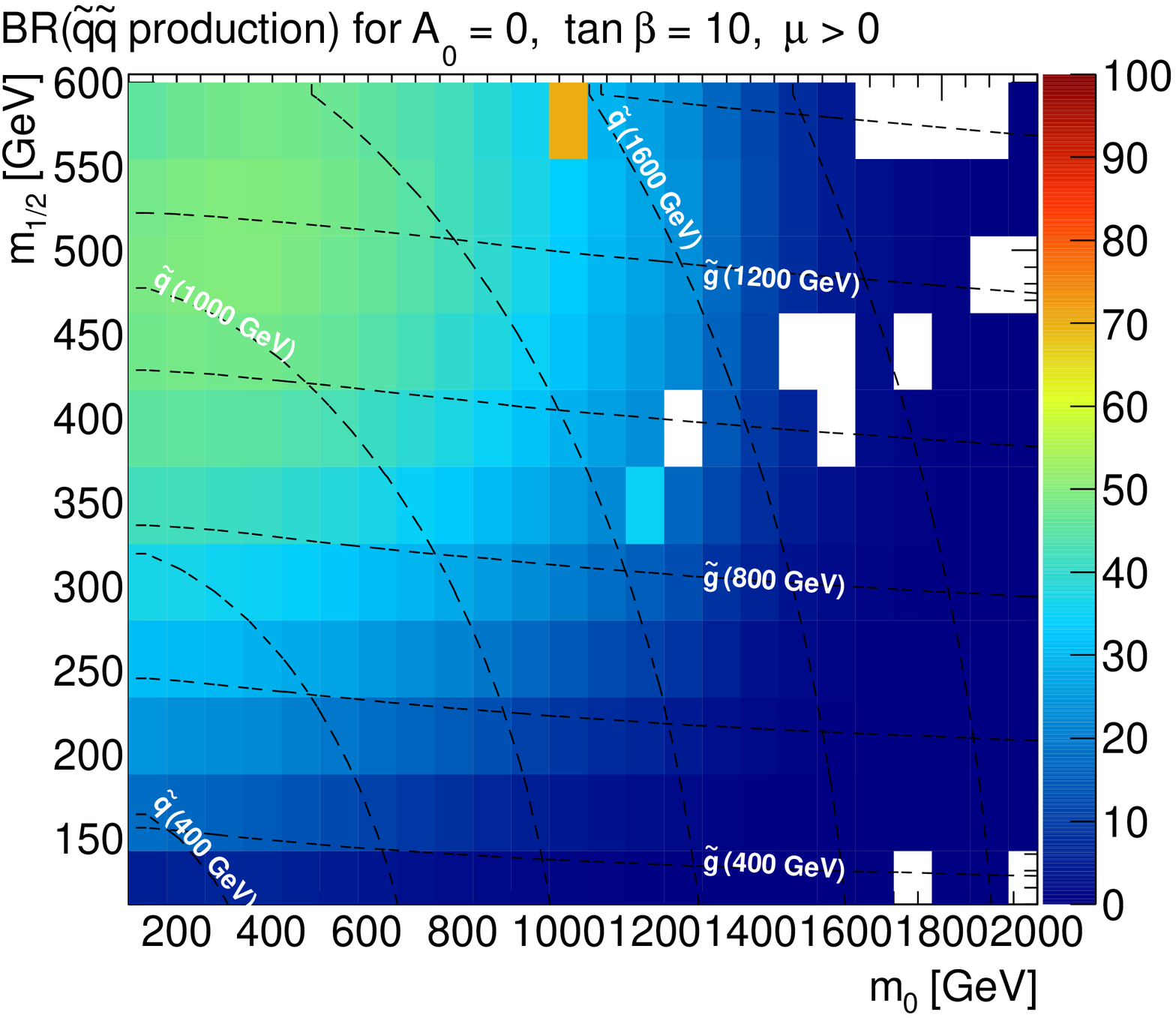}}
\quad\subfloat[$\tilde{g}\tilde{g}$-production]{\label{fig:lo2d-gl-prod}
\includegraphics[width=0.47\textwidth]{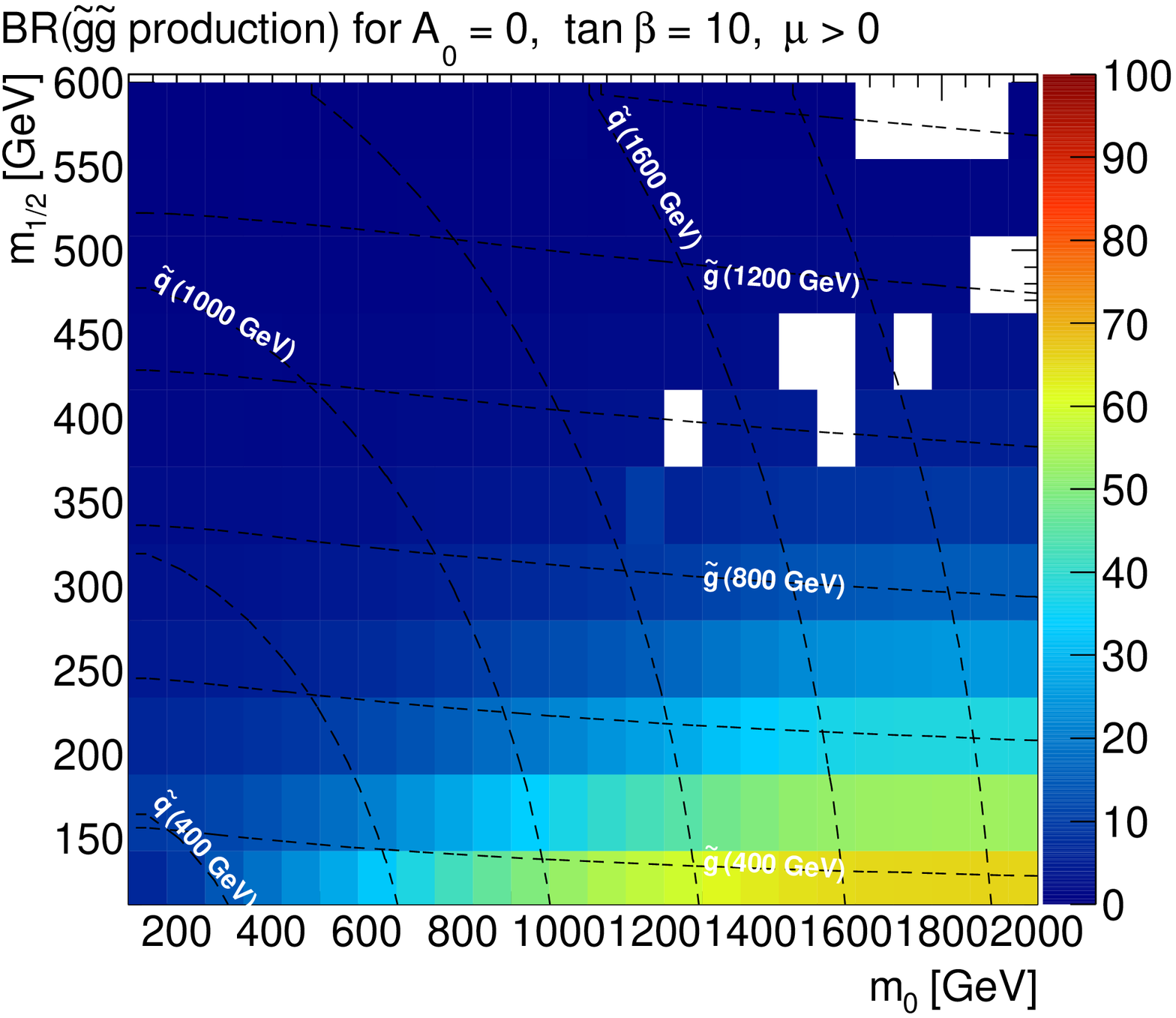}}
\linebreak\subfloat[Direct decays: $\tilde{q}\rightarrow\tilde{\chi}_1^0q\,{\rm (+jets)}$]{\label{fig:lo2d-g1-dec}
\includegraphics[width=0.47\textwidth]{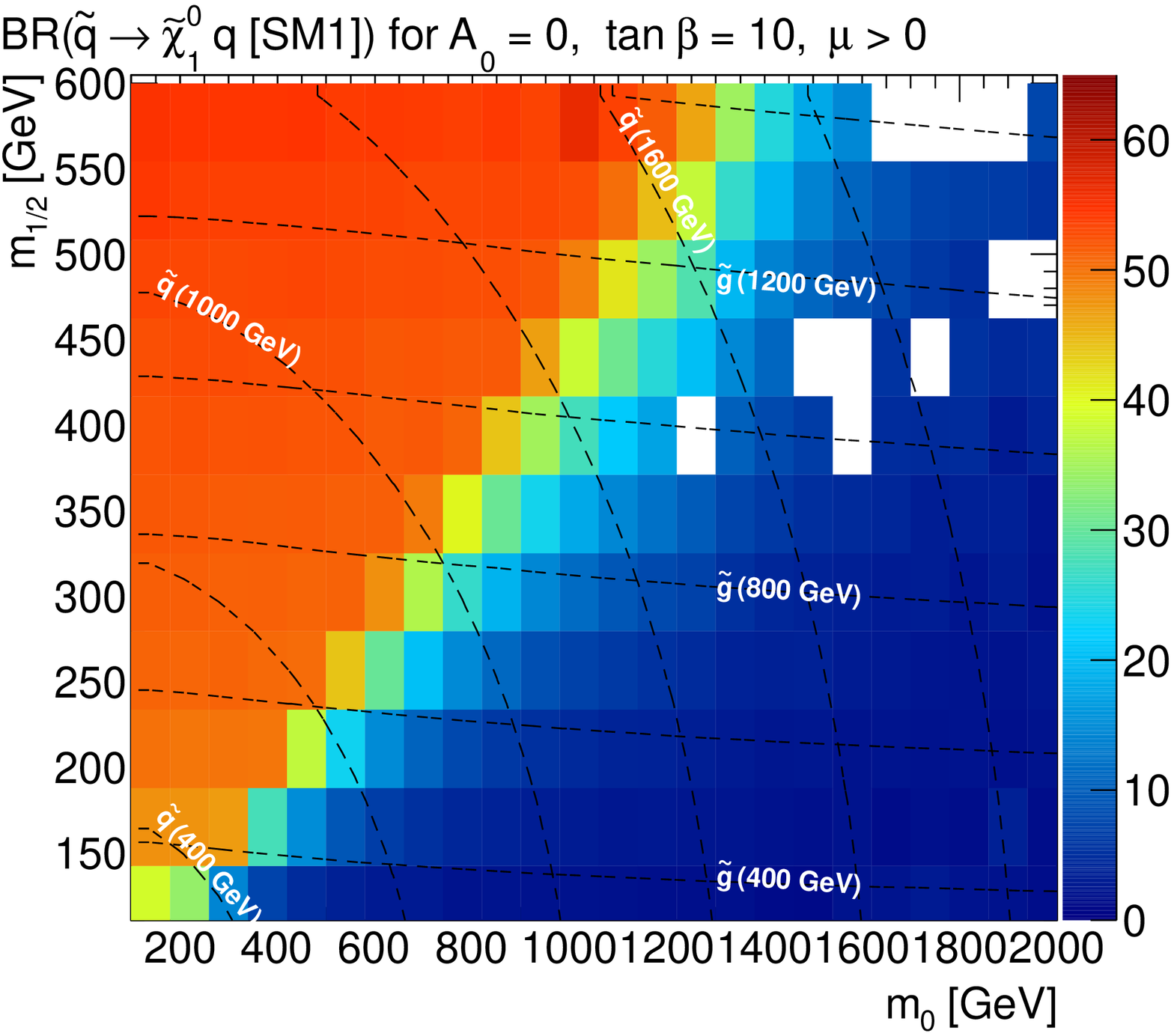}}
\quad\subfloat[Direct decays: $\tilde{g}\rightarrow\tilde{\chi}_1^0qq\,{\rm (+jets)}$]{\label{fig:lo2d-g2-dec}
\includegraphics[width=0.47\textwidth]{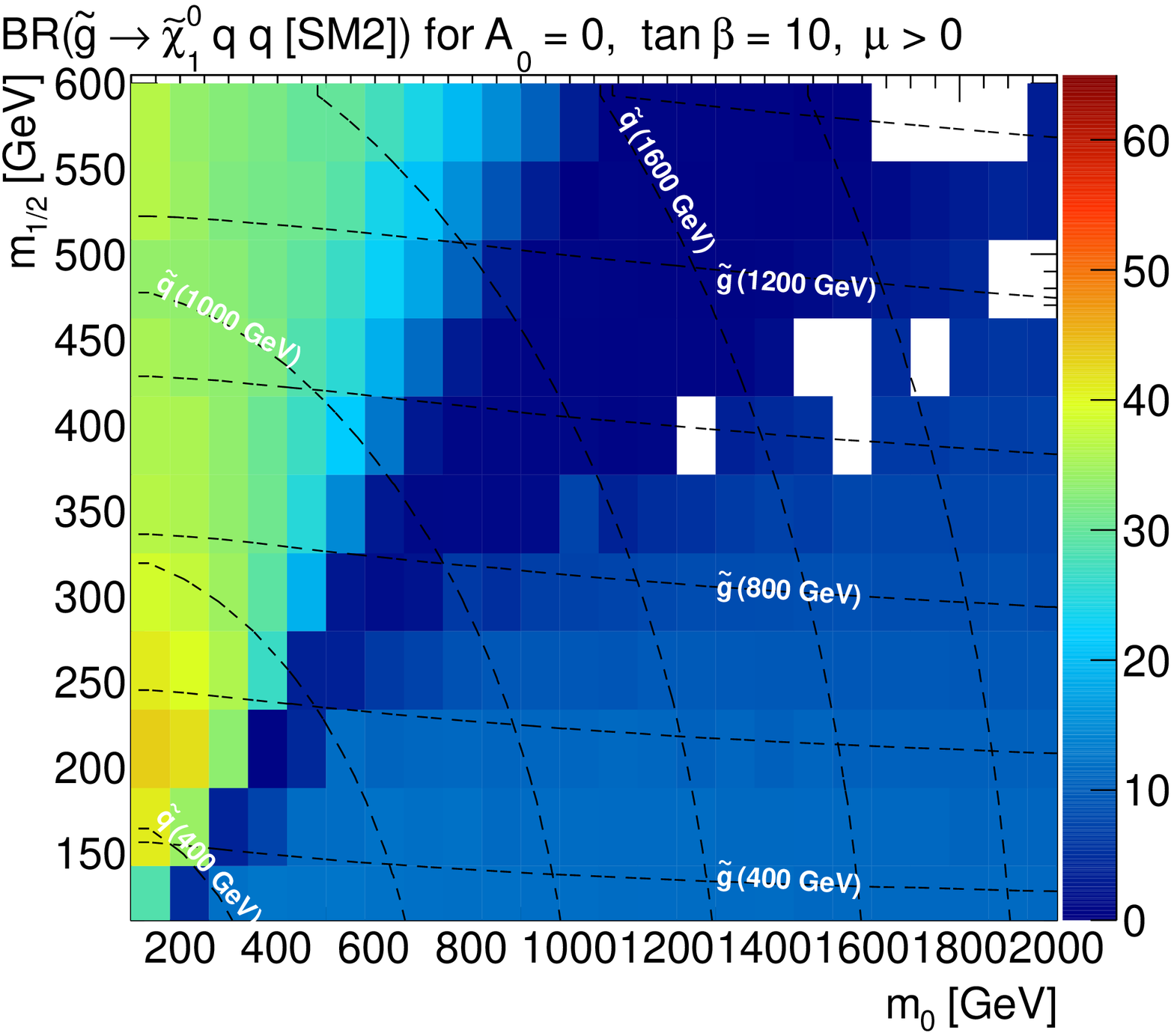}}
\caption{Variation of the branching ratios, in percent, of the main SUSY production and decay modes in the mSUGRA parameter space with $\tan\beta = 10$, $A_0 = 0$ and $\mu>0$.  The upper right corner, where the strong sparticles are heavy, includes a significant contribution from weakino production.}
\label{fig:lo2d-plots}
\end{figure*}

Despite the complexity of each point, some trends across the mSUGRA phase space are present, as demonstrated in Figure~\ref{fig:lo2d-plots}.  Squark production dominates in the low-$m_0$, high-$m_{1/2}$ region, and gluino production dominates in the high-$m_0$, low-$m_{1/2}$ region.  In the region where squark production dominates, the squark must be the lightest colored supersymmetric particle and tends to be lighter than most charginos and neutralinos. Thus, this region favors direct squark decays to the lightest supersymmetric particle (LSP).  When the gluino is light, however, the other supersymmetric particles tend to also be light, producing a large number of open decay modes and a rather more complicated topological picture.  Even when gluino production dominates, direct decays of the gluino to the LSP never comprise more than $\sim$30\,\% of the total decay phase space. These trends hold in mSUGRA and are not applicable to the full SUSY parameter space.  However, they will aid in the understanding of the limits set in Section~\ref{sec:limits}.

%
%
%
%
%

\subsubsection{Model Reconstruction}
\label{sec:reconstruction}
Five simplified model (SM) scenarios are considered for the reconstruction of these complex models, proposed in Ref.~\cite{sidewalk}:

\begin{itemize}
\item Pair-production of squarks, which directly decay to the LSP via the emission of a quark (SM 1);
\item Pair-production of gluinos, which directly decay to the LSP via the emission of a two quarks (SM 2);
\item Pair-production of squarks, which decay in one-step to the LSP.  The squark decays to a chargino via the emission of a quark, and the chargino decays to the LSP via emission of a $W$-boson (SM 3);
\item Pair-production of gluinos, which decay in one-step to the LSP.  The gluino decays to a chargino via the emission of two quarks, and the chargino decays to the LSP via emission of a $W$-boson (SM 4); and
\item Pair-production of charginos, which directly decay to the LSP via the emission of a $W$-boson (SM 5).
\end{itemize}

\begin{figure*}
\centering
\graphicspath{{plots/}}
\subfloat{
\includegraphics[width=0.47\textwidth]{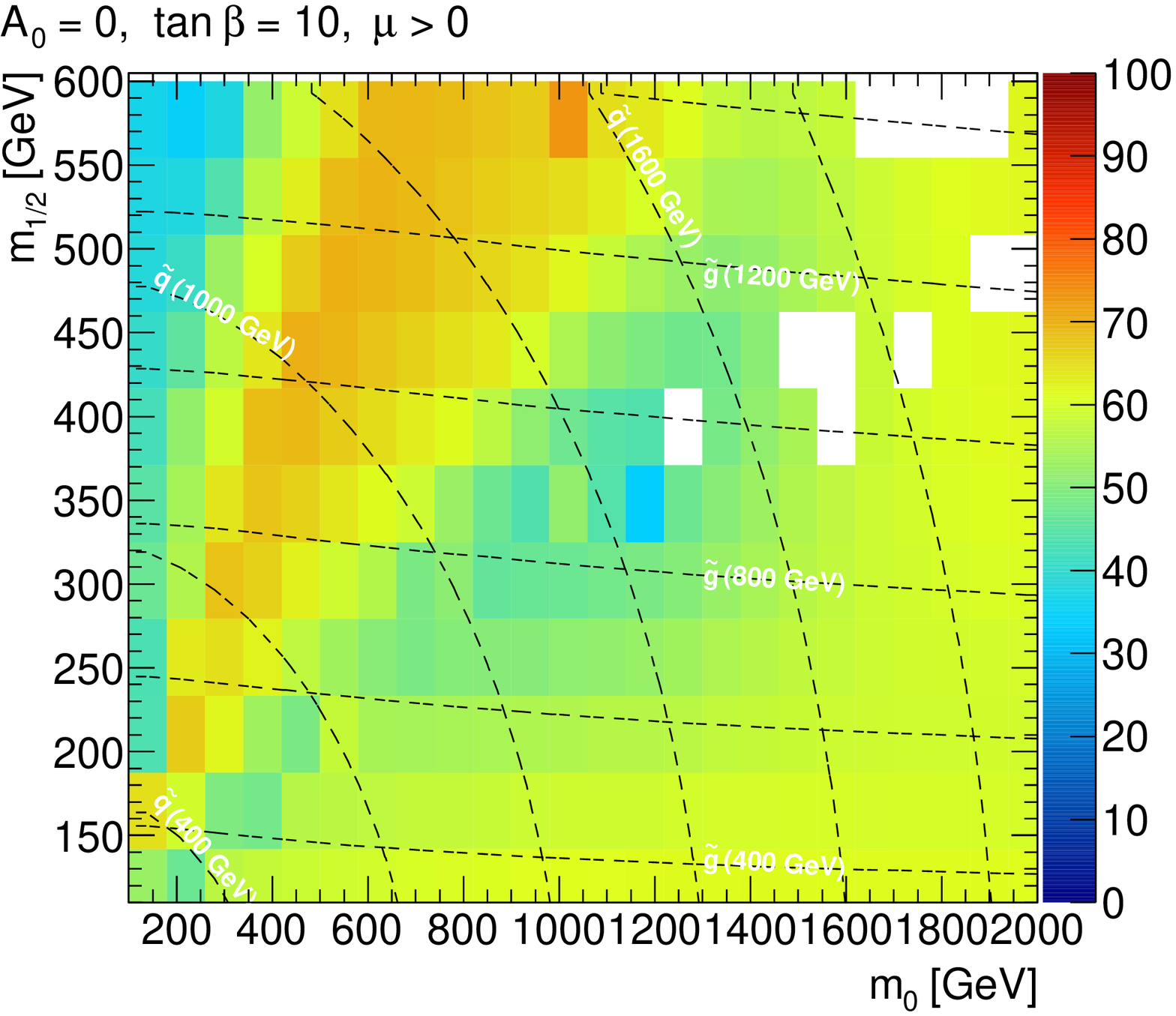}}
\quad\subfloat{
\includegraphics[width=0.47\textwidth]{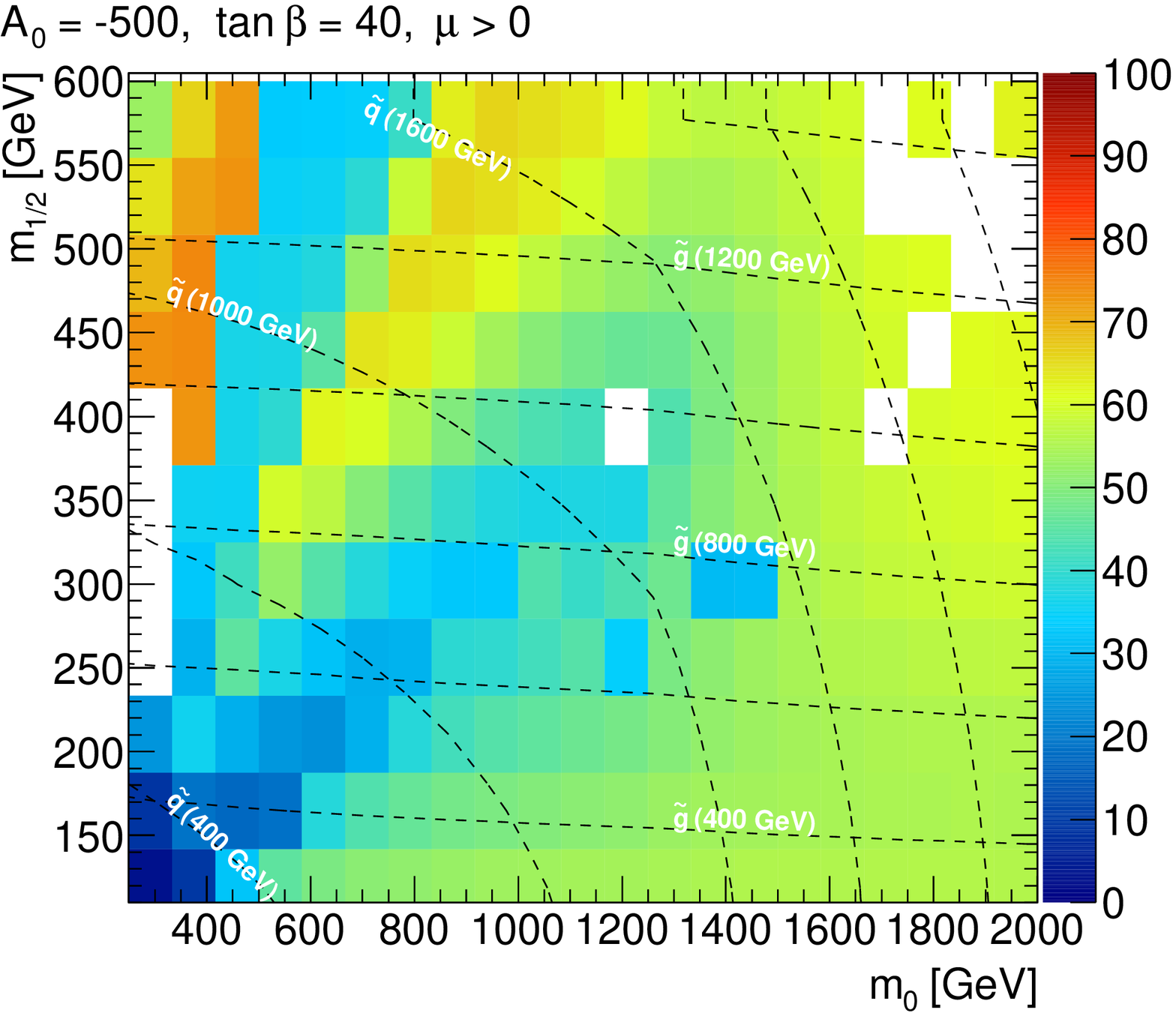}}
\caption{
The fraction of mSUGRA events classified as belonging to one of the five simplified models considered in this paper, for low-$\tan{\beta}$ (left) and high-$\tan{\beta}$ (right).
}
\label{fig:smclassification}
\end{figure*}

In order to reconstruct a complex mSUGRA point from simplified models, only the masses, production fractions, and branching fractions are needed.  When the squark decays to the gluino, the gluino decay is counted in classifying the event topology, and the decay of the squark to the gluino is counted as an additional jet in the event (``plus jets''), as though it were identical to initial- or final-state radiation.   
When the gluino decays through a squark ($\tilde{g} \rightarrow q \tilde{q}$, $\tilde{q} \rightarrow q \text{X}$), however, the final state of the decay still appears as though the gluino had produced two jets and decayed directly, omitting the squark-step, save some (small) differences in kinematics.  For these cases, therefore, the decay chain is classified as though the gluino decayed via the emission of a pair of quarks with no intermediate squark ($\tilde{g} \rightarrow q q \text{X}$), rather than classifying it as the squark decay with an additional initial- or final-state radiation-like jet ($\tilde{q} \rightarrow q \text{X}$ plus jet(s)).  Associated squark-gluino production is divided evenly among the squark and gluino simplified models.  With these approximations, it is possible to classify a large fraction of SUSY events as one of the five simplified models under consideration.  The fraction of SUSY events classified as belonging to one of these five simplified models is shown in Figure~\ref{fig:smclassification}.

\begin{figure*}
\centering
\graphicspath{{plots/}}
\subfloat{
\includegraphics[width=0.47\textwidth]{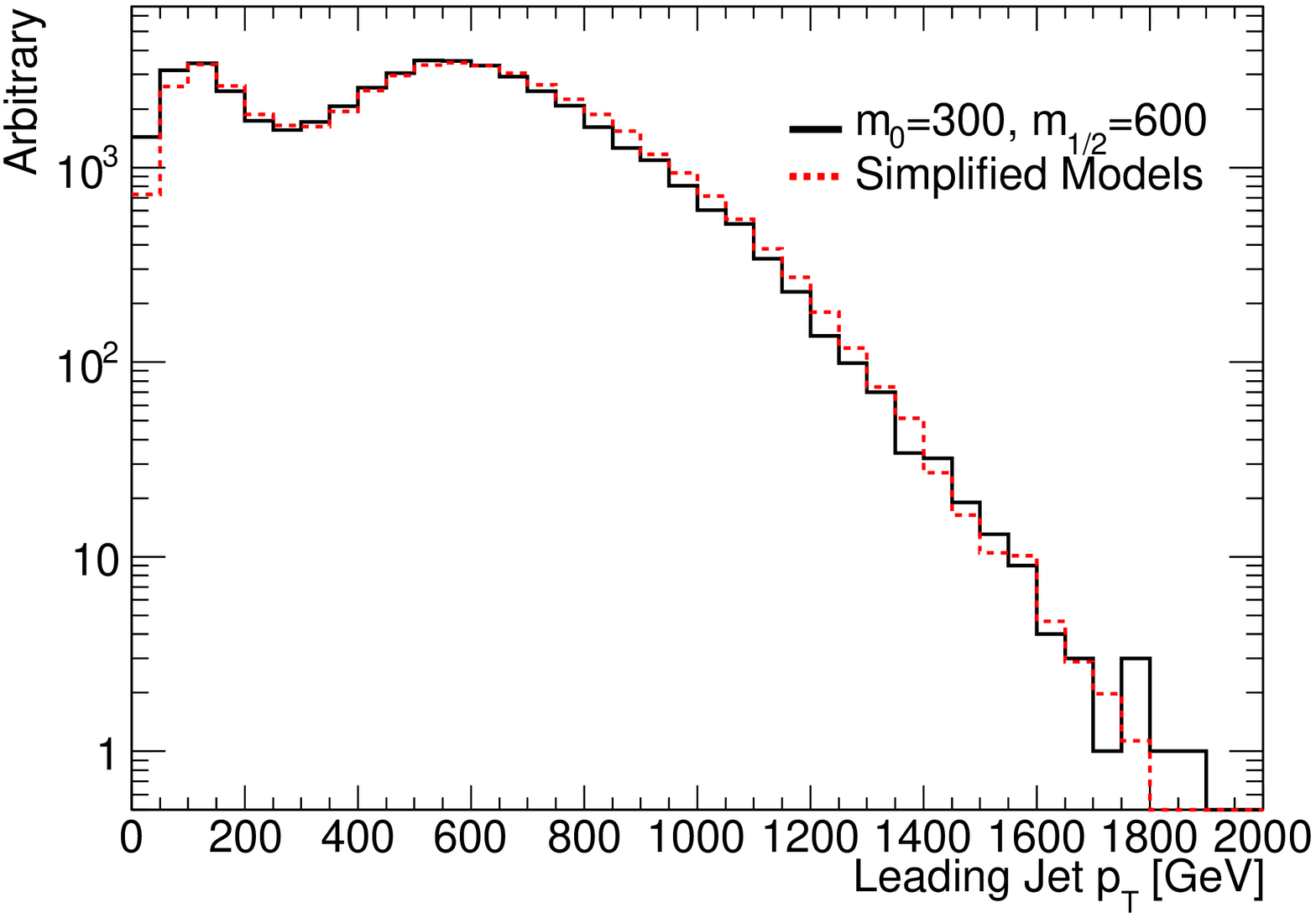}}
\quad{
\includegraphics[width=0.47\textwidth]{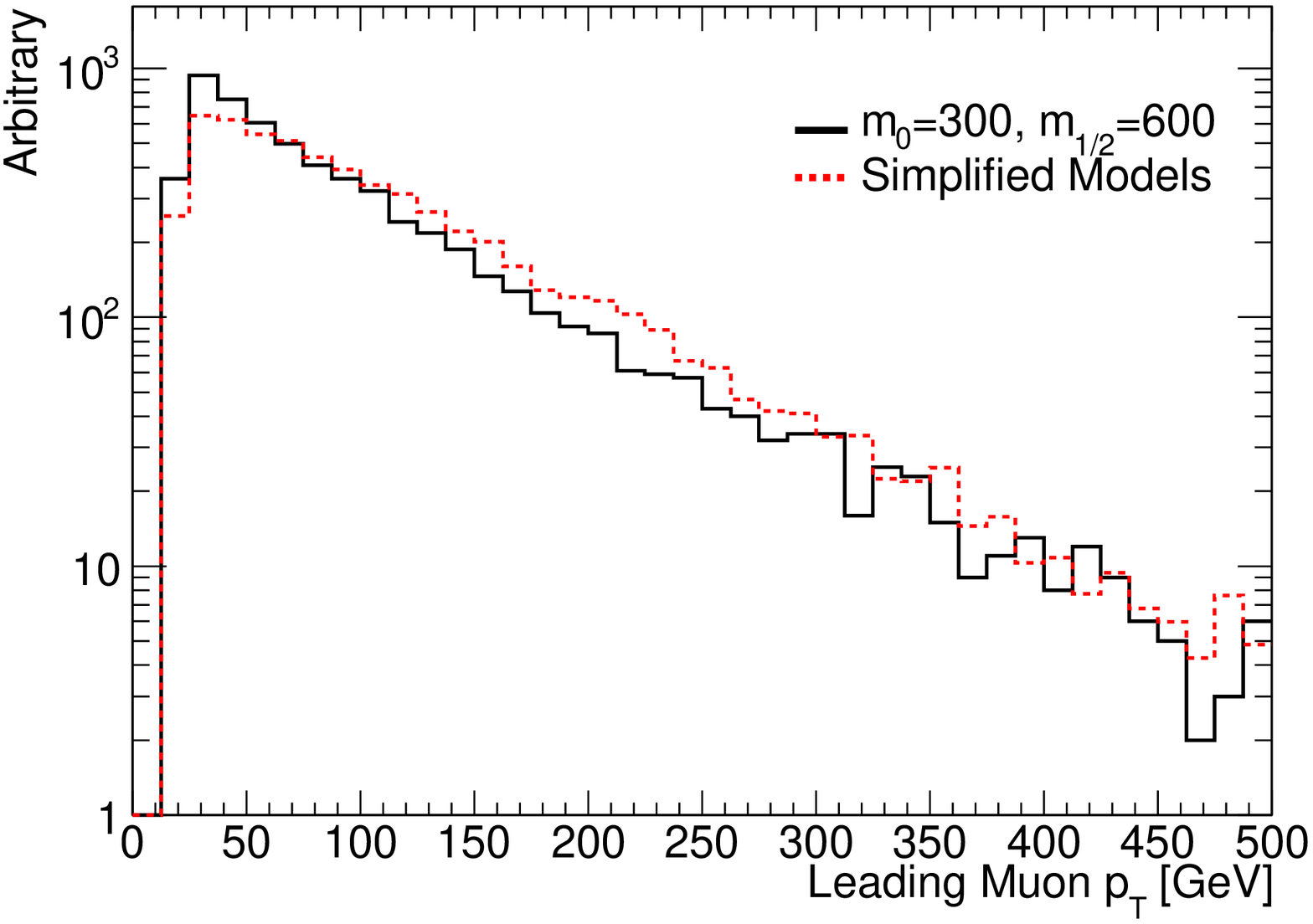}}
\quad{
\includegraphics[width=0.47\textwidth]{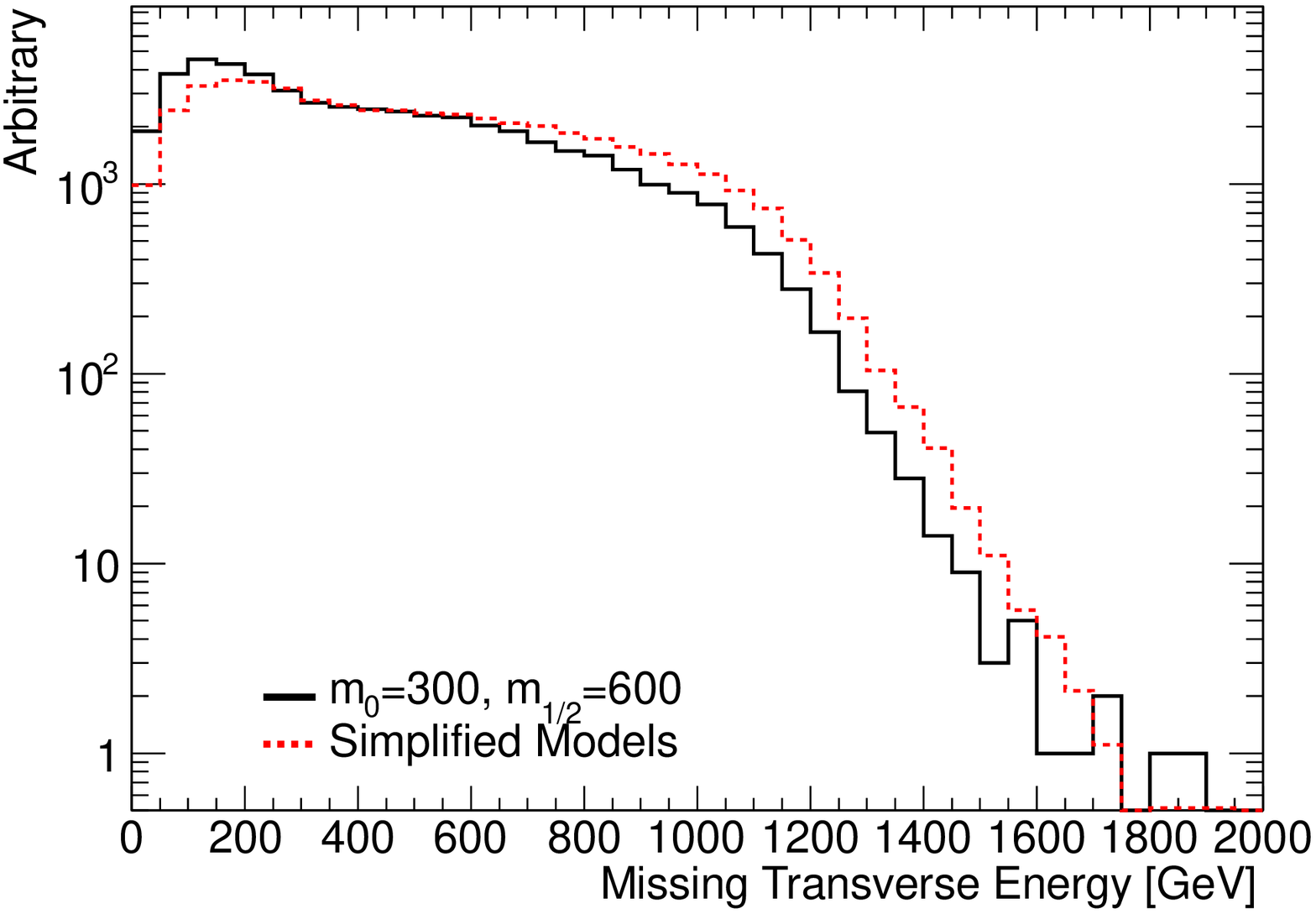}}
\quad{
\includegraphics[width=0.47\textwidth]{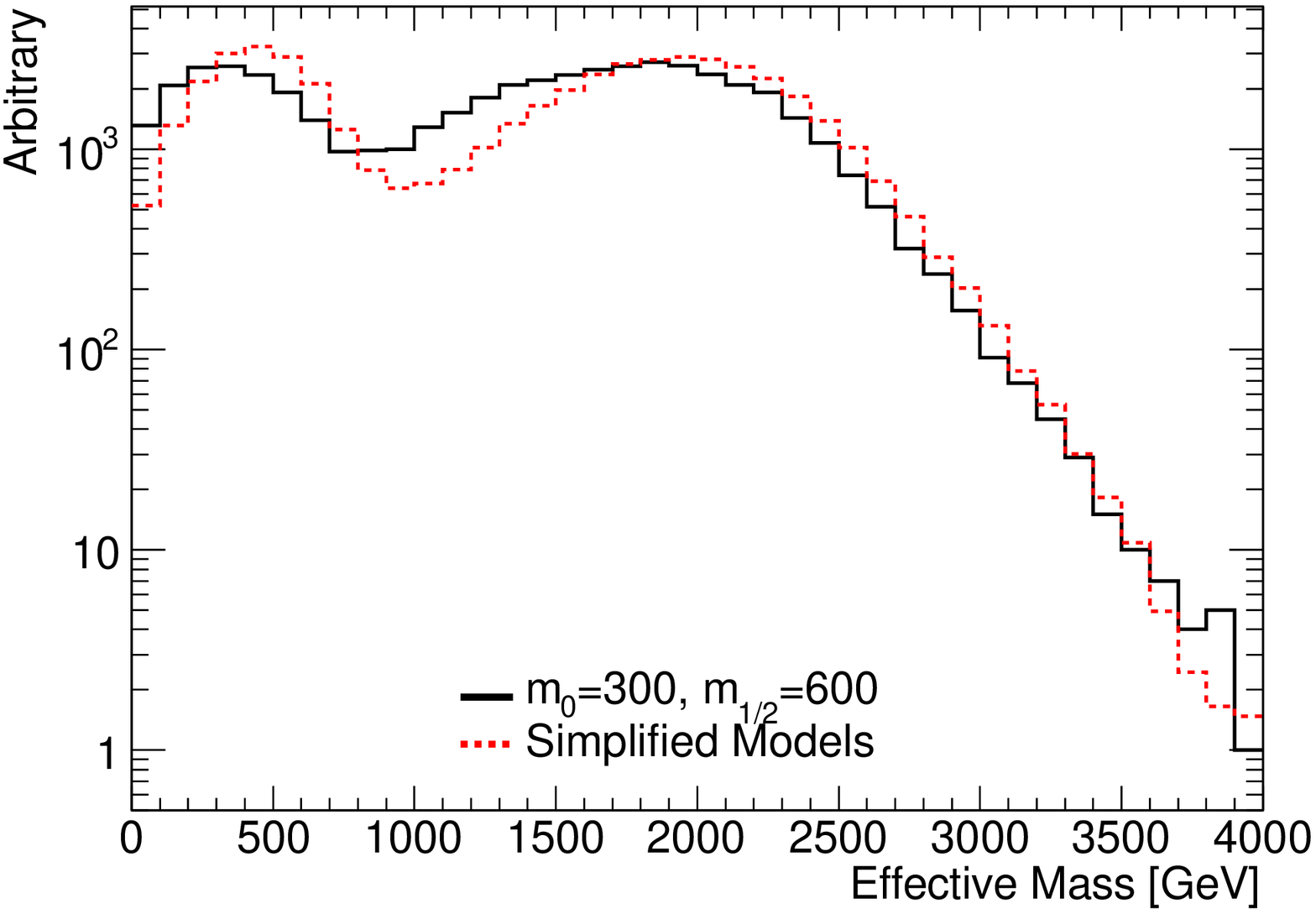}}
\caption{
Kinematics of a squark-production-dominated mSUGRA point ($m_0=300$~GeV, $m_{1/2}=600$~GeV, $\tan(\beta)=10$, $A_0=0$, $\mu>0$) and a set of five simplified models constructed using the same mass spectrum.  Clockwise from the top left, leading jet $p_T$, leading muon $p_T$, effective mass, and missing transverse energy.  No signal selection has been applied.  
}
\label{fig:kinematics1}
\end{figure*}

\begin{figure*}[tb]
\centering
\graphicspath{{plots/}}
\includegraphics[width=0.47\textwidth]{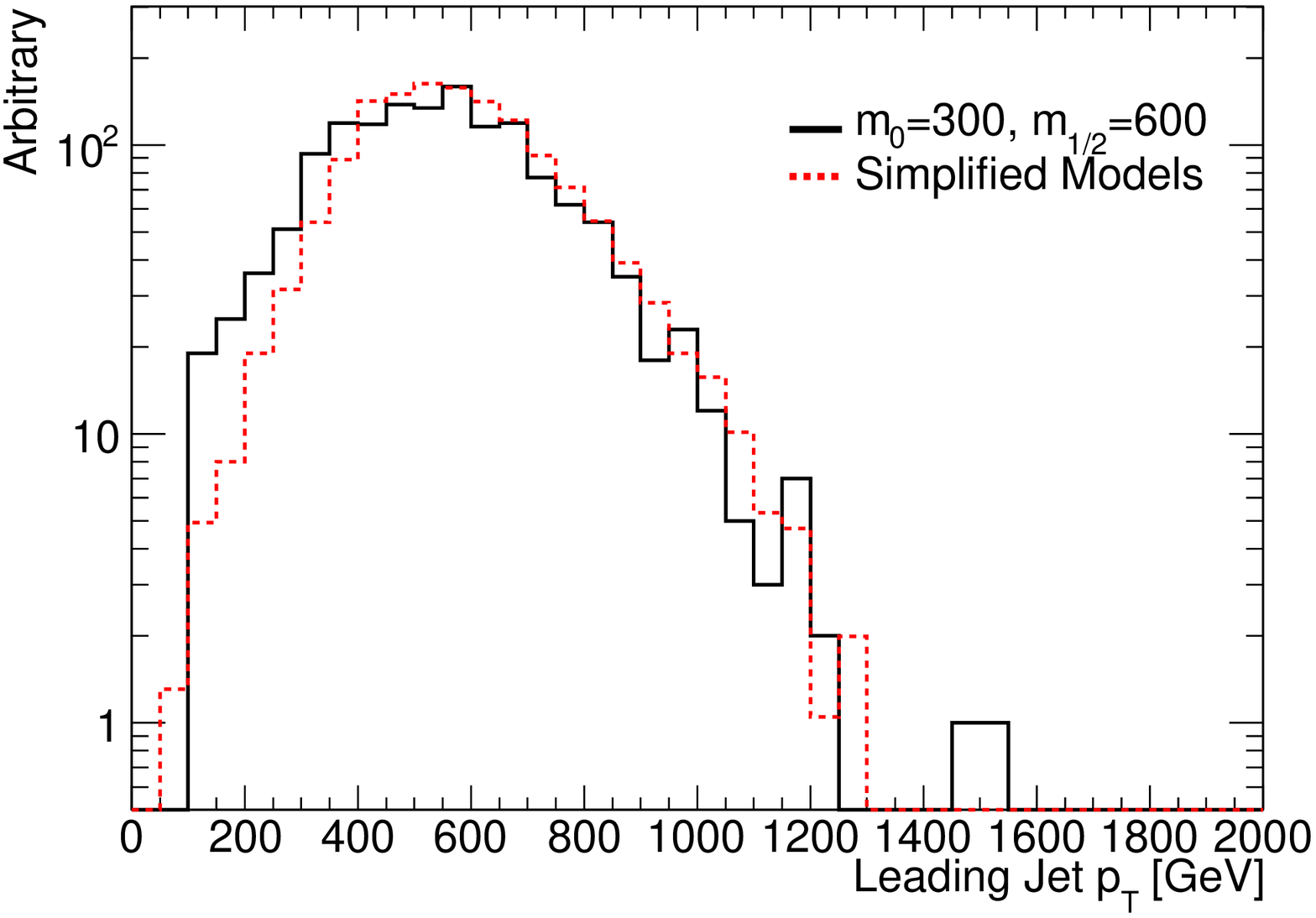} 
\includegraphics[width=0.47\textwidth]{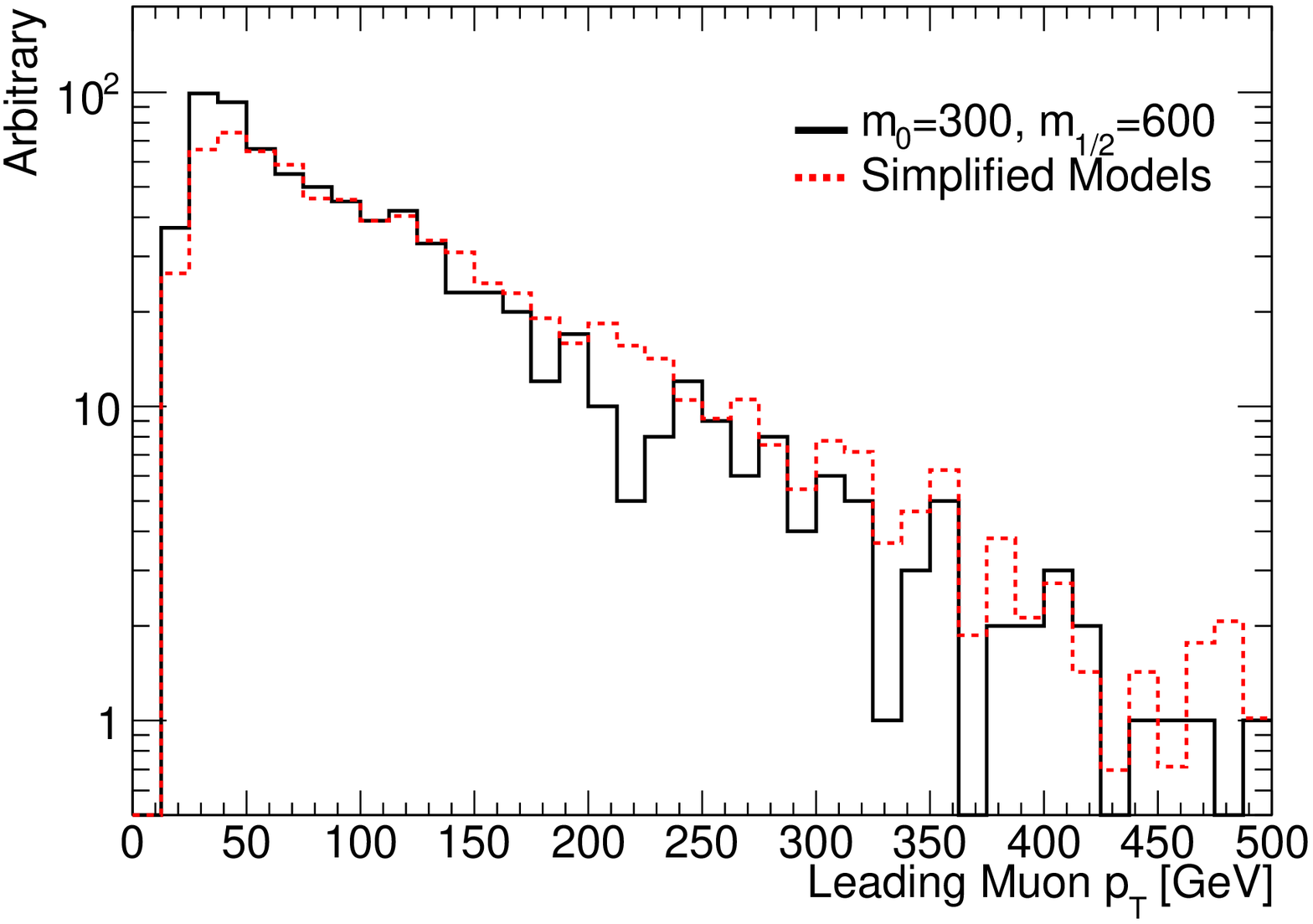} 
\includegraphics[width=0.47\textwidth]{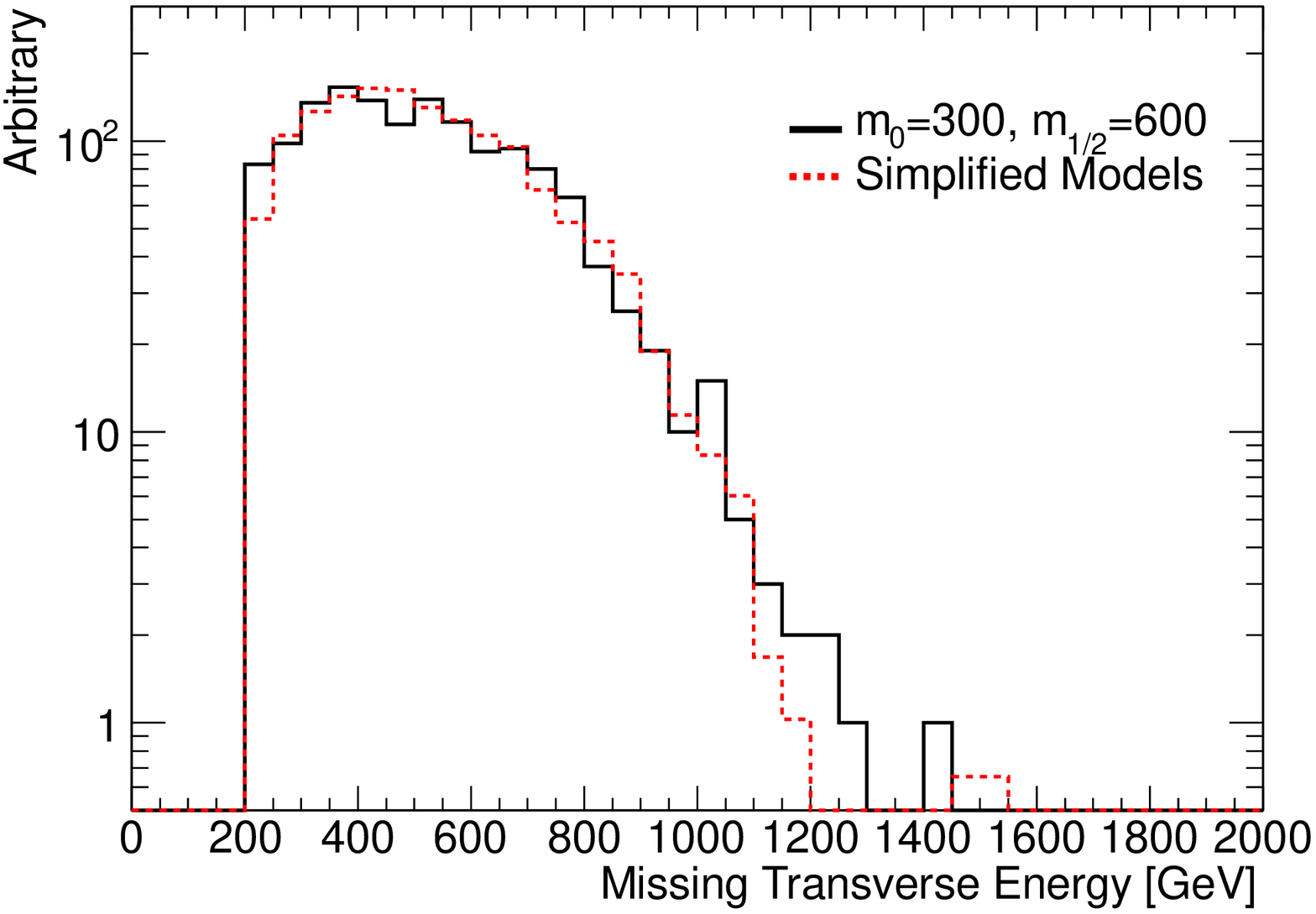} 
\includegraphics[width=0.47\textwidth]{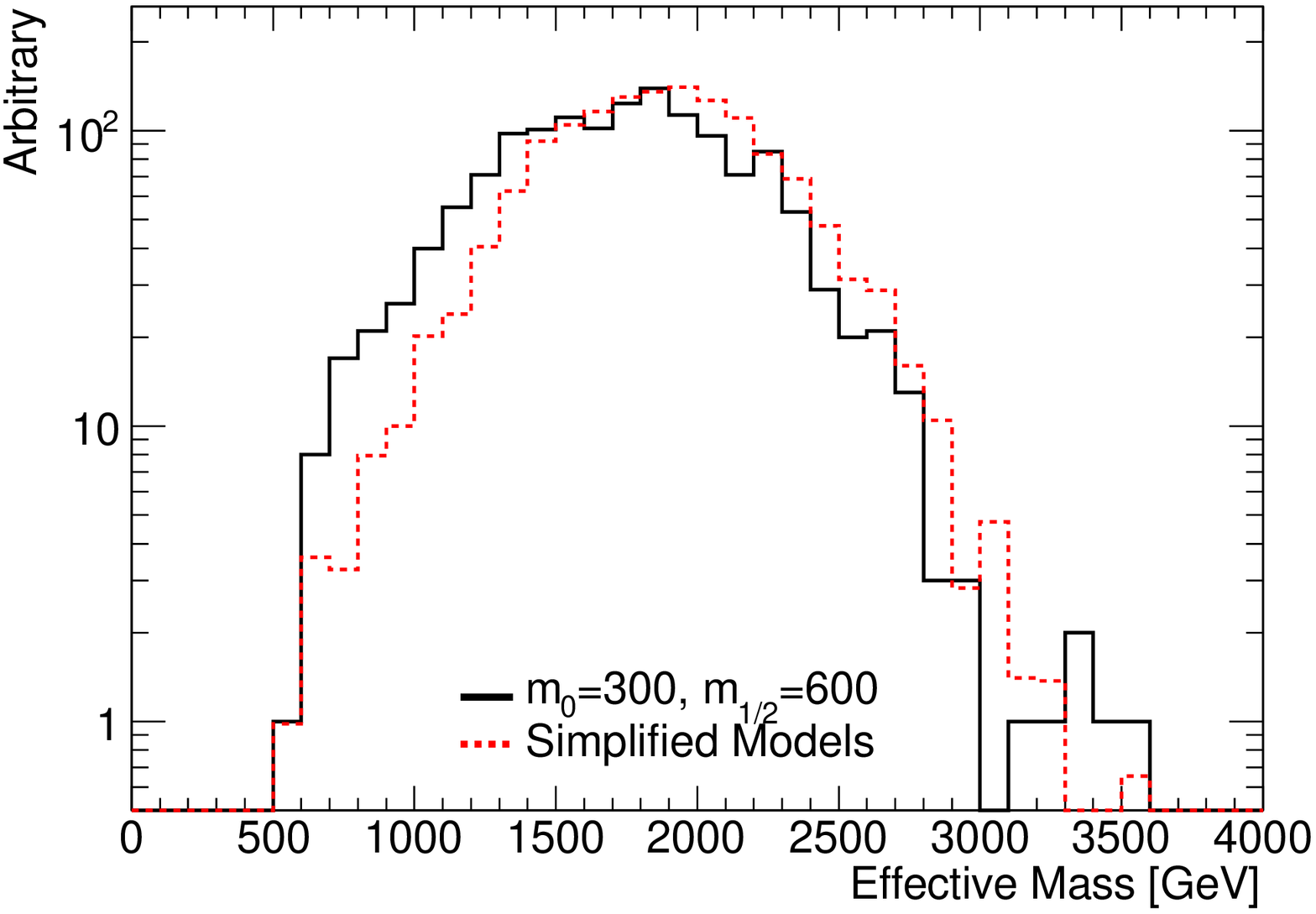} 
\caption{
Kinematics of a squark-production-dominated mSUGRA point ($m_0=300$~GeV, $m_{1/2}=600$~GeV, $\tan(\beta)=10$, $A_0=0$, $\mu>0$) and a set of five simplified models constructed using the same mass spectrum.  Clockwise from the top left, leading jet $p_T$, leading muon $p_T$, effective mass, and missing transverse energy.  A signal selection similar to the one-lepton four-jet ``tight'' ATLAS SUSY search has been applied.  
}
\label{fig:kinematics2}
\end{figure*}

\begin{figure*}[tb]
\centering
\graphicspath{{plots/}}
\includegraphics[width=0.47\textwidth]{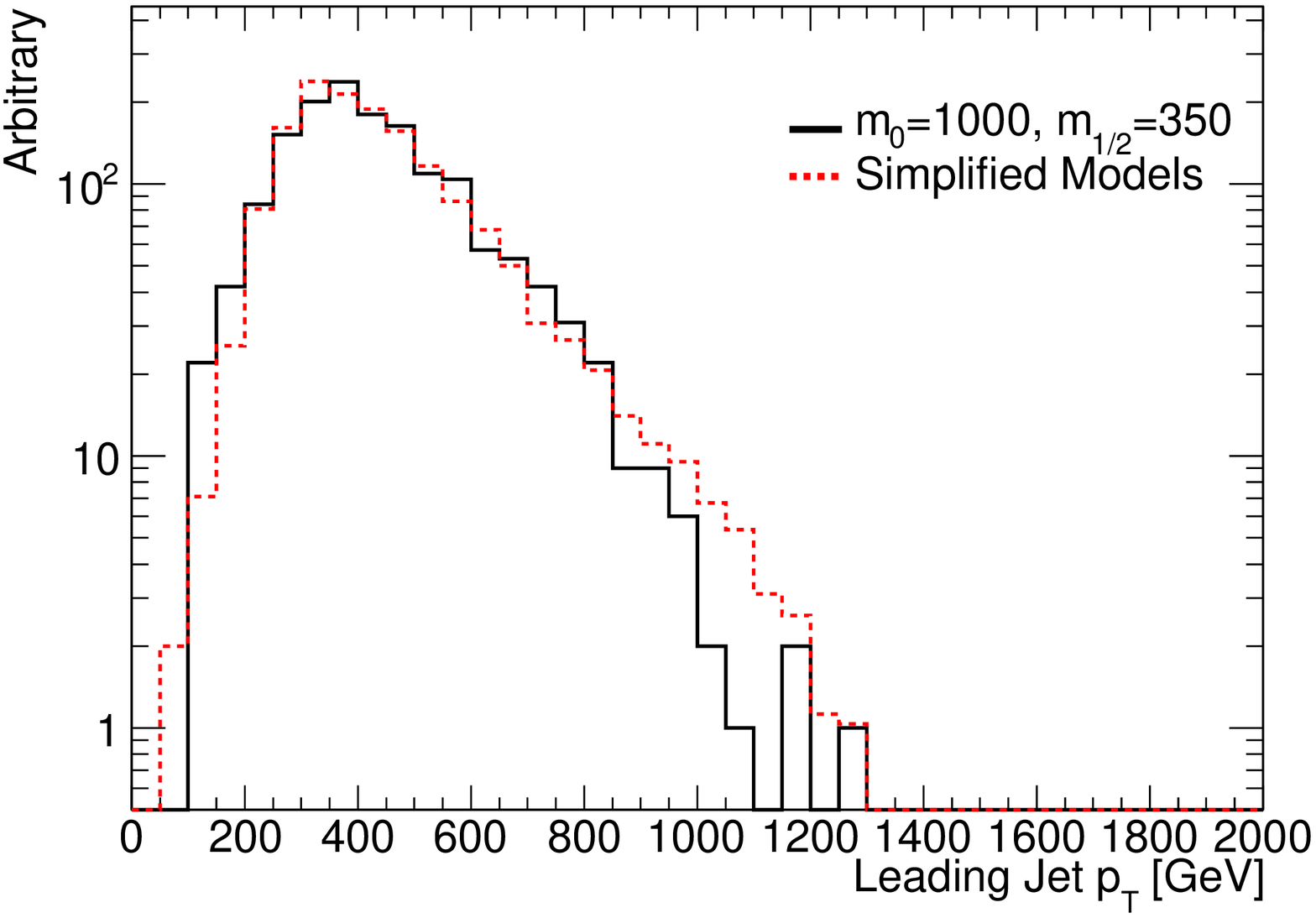} 
\includegraphics[width=0.47\textwidth]{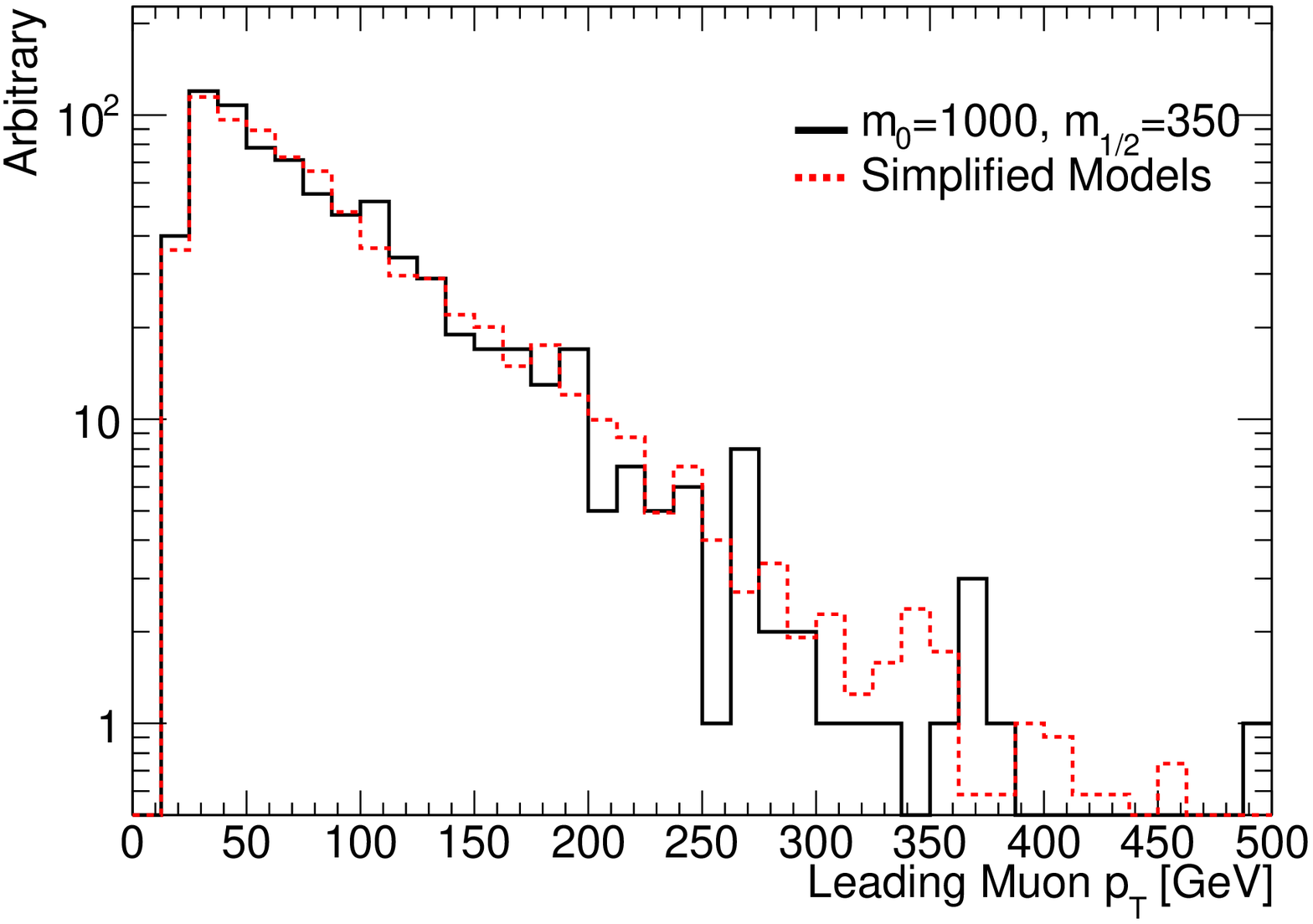} 
\includegraphics[width=0.47\textwidth]{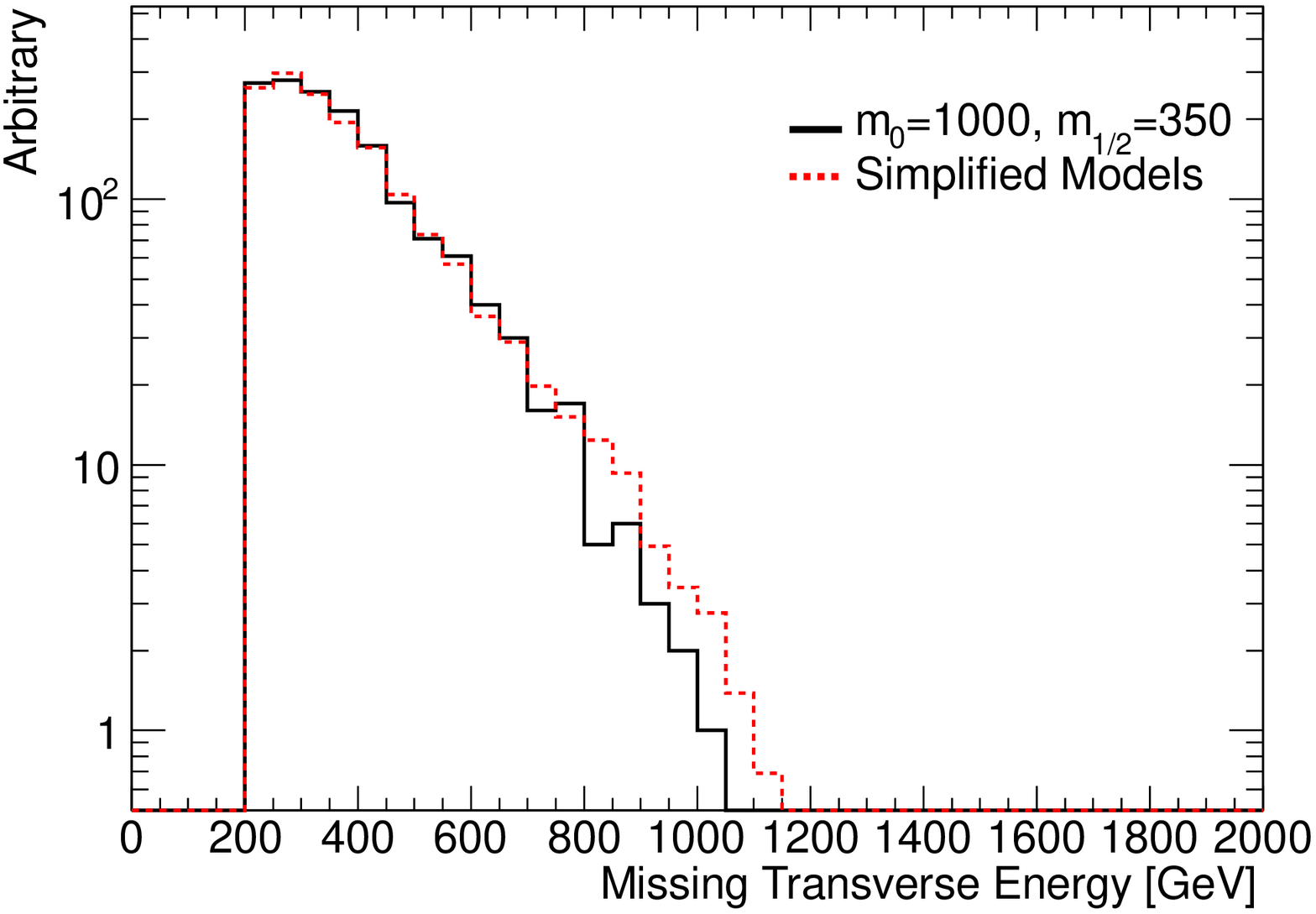} 
\includegraphics[width=0.47\textwidth]{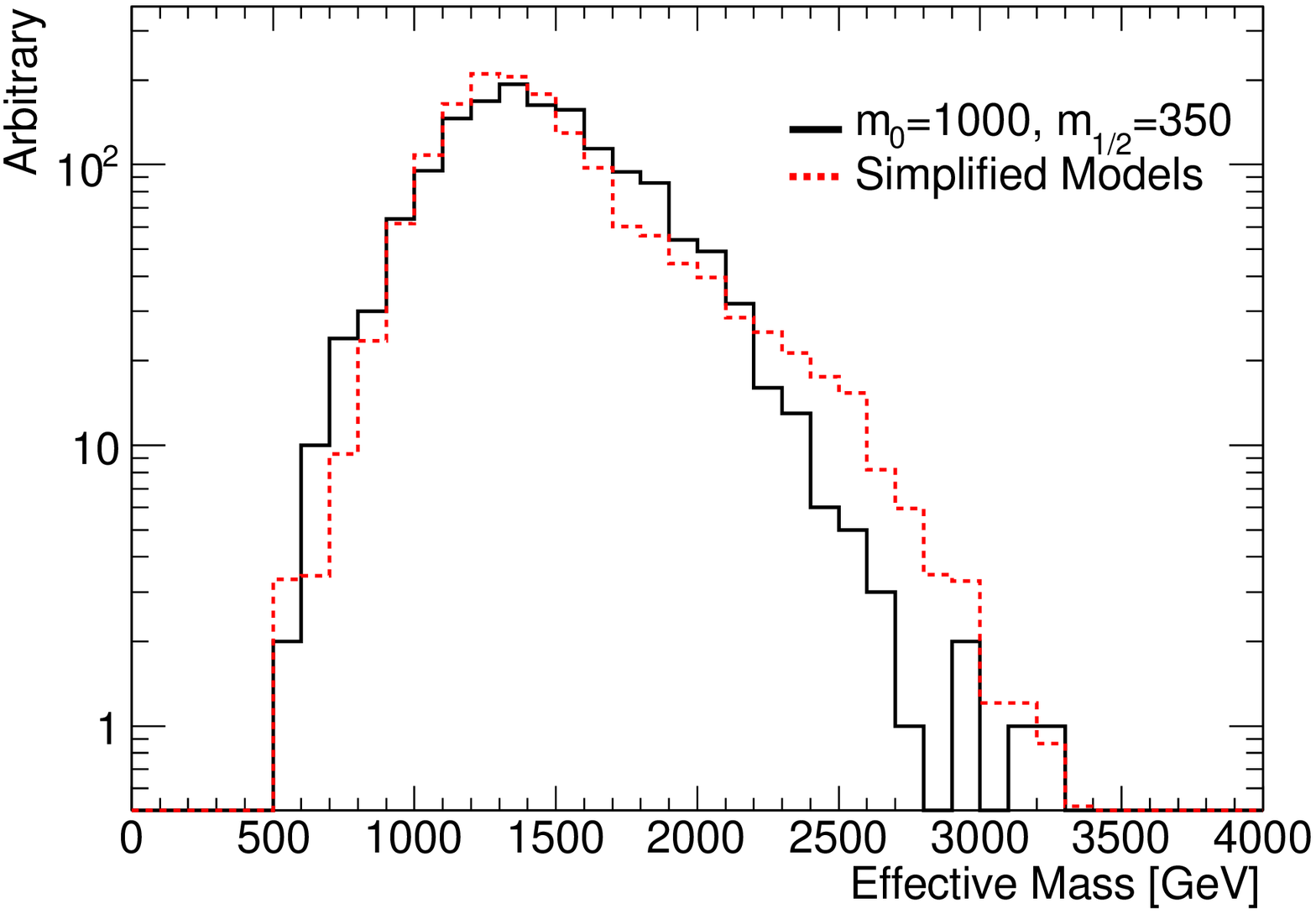} 
\caption{
Kinematics of a complex mSUGRA point ($m_0=1000$~GeV, $m_{1/2}=350$~GeV, $\tan(\beta)=10$, $A_0=0$, $\mu>0$) and a set of five simplified models constructed using the same mass spectrum.  Clockwise from the top left, leading jet $p_T$, leading muon $p_T$, effective mass, and missing transverse energy.  A signal selection similar to the one-lepton four-jet ``tight'' ATLAS SUSY search has been applied.  
\label{fig:kinematics3}
}
\end{figure*}

The event kinematics for two mSUGRA parameter space points, along with a combination of simplified models used to mimic them, are shown in Figures~\ref{fig:kinematics1}, \ref{fig:kinematics2}, and~\ref{fig:kinematics3}.  These two points are deconstructed using the method described above, and five simplified models are constructed and combined according to the mass spectra, production rates, and branching fractions of the points.  The simplified model events were generated and analyzed in a manner identical to the mSUGRA events.  Here, four of the key kinematic variables used in LHC supersymmetry searches are shown: leading jet transverse momentum ($p_T$), lepton $p_T$, missing transverse energy, and effective mass, defined as the scalar sum of the transverse momenta of the four leading jets and the lepton.  Two features are visible in the effective mass, leading jet, and missing transverse energy distributions, corresponding to strong production and weakino production.  Some discrepancies are clearly visible.  The low-$p_T$ lepton tail, for example, is predominantly from tau decays that are not covered by any of the simplified models.  The low missing transverse energy, low effective mass region is in part from LSP-X associated production, which is not modeled.  

However, the cuts of most signal regions used at the LHC are such that simple decay topologies are selected over the more complex, often softer or higher multiplicity events.  Thus, signal region selection tends to improve the description of event kinematics by simplified models.  Comparison in a one-lepton region similar to that used in a recent ATLAS SUSY search~\cite{0lpaper} are shown in Figure~\ref{fig:kinematics2} and \ref{fig:kinematics3}.  The agreement in both shape and tails is significantly better.  The kinematics for the simplified models compare well to the {\it inclusive} SUSY model kinematics, suggesting that the efficiency and acceptance for complete SUSY point may be well described by a limited combination of simplified models.  Of course, the kinematics of only those SUSY events corresponding to topologies described by the simplified models are identical to their simplified model counterparts.  This serves as an indication that those events not covered by these simplified models are either a small fraction or kinematically similar to those that are.

\renewcommand*\thesubsubsection{\Roman{subsubsection}}
\subsubsection{Approximating Limits}
\label{sec:limits}
The deconstruction of mSUGRA points into components and the application of simplified model limits to those components allows the construction of an mSUGRA exclusion limit directly.  For any signal region, the most conservative limit can be set using the production fraction $P_{(a,a)}$ (where $a$ represents some sparticle and all five simplified models considered here include pair production) of events identical to a simplified model $i$ and the branching fraction for the produced sparticle to decay in the manner described by the simplified model, $\text{BR}_{a\rightarrow i}$.  The expected number of events in a given signal region from these simple topologies can then be written as

\begin{equation}
N = \sigma_\text{tot} \times \mathcal{L}_\text{int} \times \sum_\text{SM} A\epsilon_{a\rightarrow i} \times P_{(a,a)} \times \text{BR}_{a\rightarrow i}^2,
\label{eq:first}
\end{equation}

\noindent where the sum is over simplified models, $\sigma_\text{tot}$ is the total cross section for the mSUGRA point, $\mathcal{L}_\text{int}$ is the integrated luminosity used in the search, and $A\epsilon_{a\rightarrow i}$ is the acceptance times efficiency for the simplified model events in the signal region being considered.  This number can be compared to the expected 95\,\% confidence level upper limit on the number of new physics events to select the optimal search region.  The model can then be excluded if $N$ is larger than the observed number of new physics events excluded at the 95\,\% confidence level.  Exclusions in non-overlapping regions could be combined if information about the correlations of their uncertainties were available.

\begin{figure*}
\centering
\graphicspath{{plots/}}
\subfloat[Combined zero-lepton SRs]{
\includegraphics[width=0.47\textwidth]{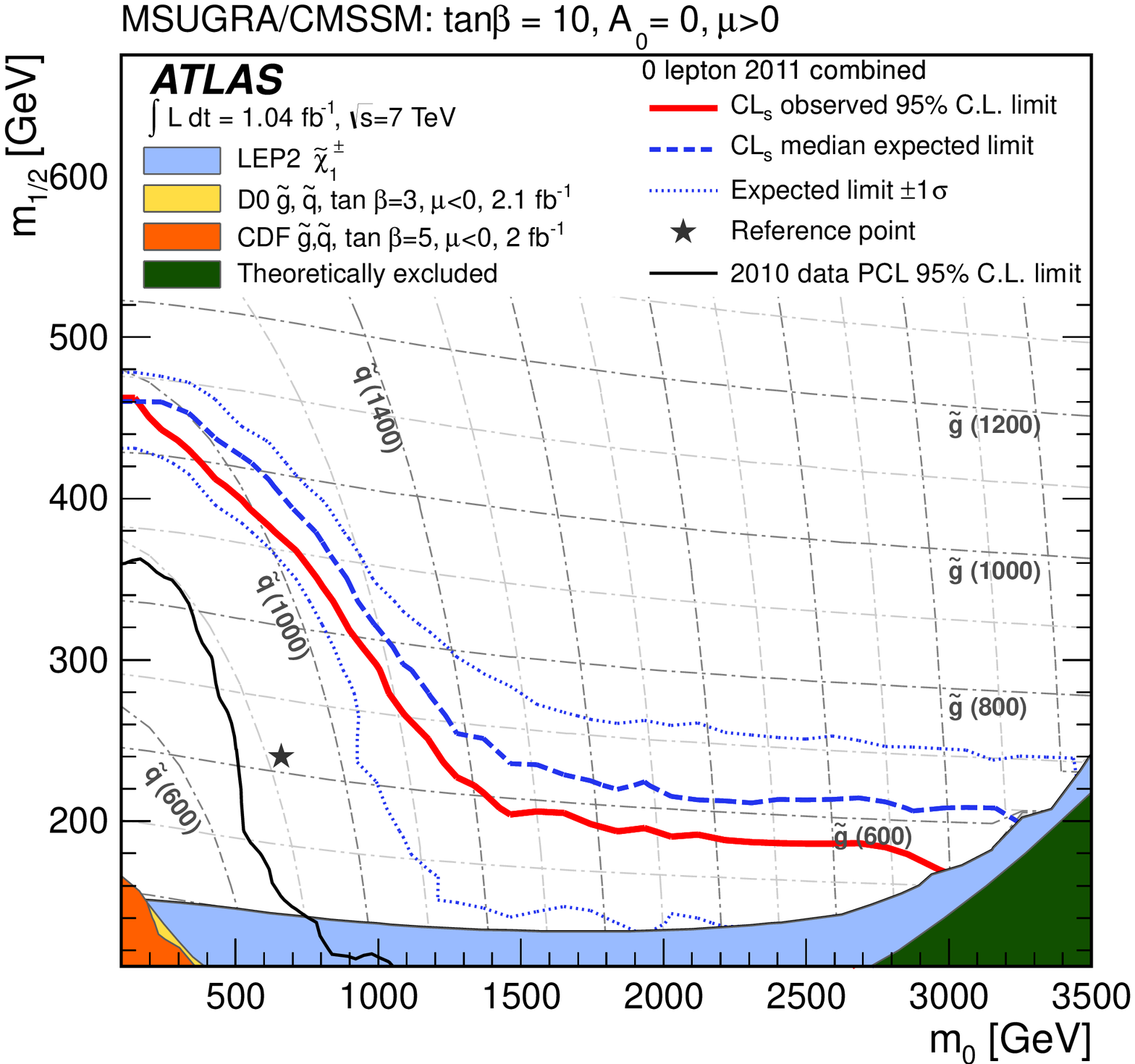}}
\quad\subfloat[Simplified models]{
\includegraphics[width=0.47\textwidth]{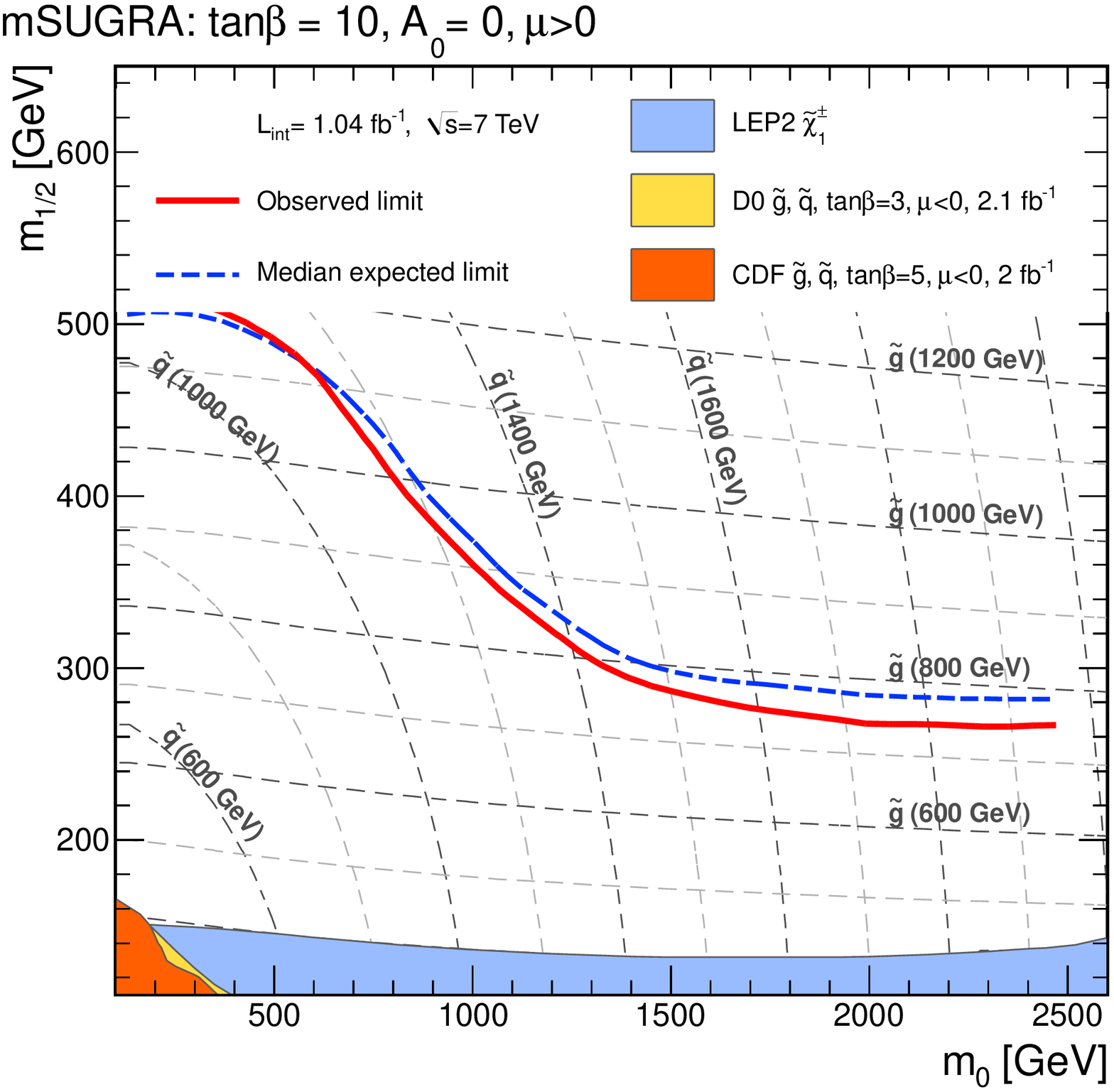}}
\caption{
Combined zero-lepton exclusion limits for mSUGRA models with $\tan\beta=10$, $A_0=0$ and $\mu>0$ from Ref.~\cite{0lpaper} (left) in comparison with the exclusion limit obtained using PGS and without a systematic uncertainty on the signal. The signal region providing the best expected limit is taken for a given point in parameter space. The expected 95\,\% confidence level limit is shown as a dashed blue line, and the observed limit is shown as a solid red line. Results from previous searches are also shown for comparison purposes \cite{d-zero-1}--\cite{atlas-2010}, although some of these limits were produced using slightly different parameter choices.
}
\label{fig:nosystlimit}
\end{figure*}

This prescription omits the treatment of theoretical uncertainties on the signal model.  In fact, the LHC experiments currently do not treat these uncertainties in a consistent way, nor are all of the uncertainties included.  No experiment, for example, includes any uncertainty in the calculation of visible masses from the GUT scale parameters.  The limits that are presented here, therefore, should be expected to differ from the published limits.  In Figure~\ref{fig:nosystlimit}, the published ATLAS exclusion limits in the zero-lepton channel are compared to those obtained here without any systematic uncertainty on the signal.  The limit without signal uncertainties is clearly higher than the published limit.  For the remainder of the paper, the limit without systematic uncertainties on the signal will be taken as the ``correct answer'' to be arrived at using simplified models.  The theoretical uncertainty can be added to both in the same way and will affect both limits in approximately the same way.

\begin{figure*}
\centering
\graphicspath{{plots/}}
\subfloat{
\includegraphics[width=0.49\textwidth]{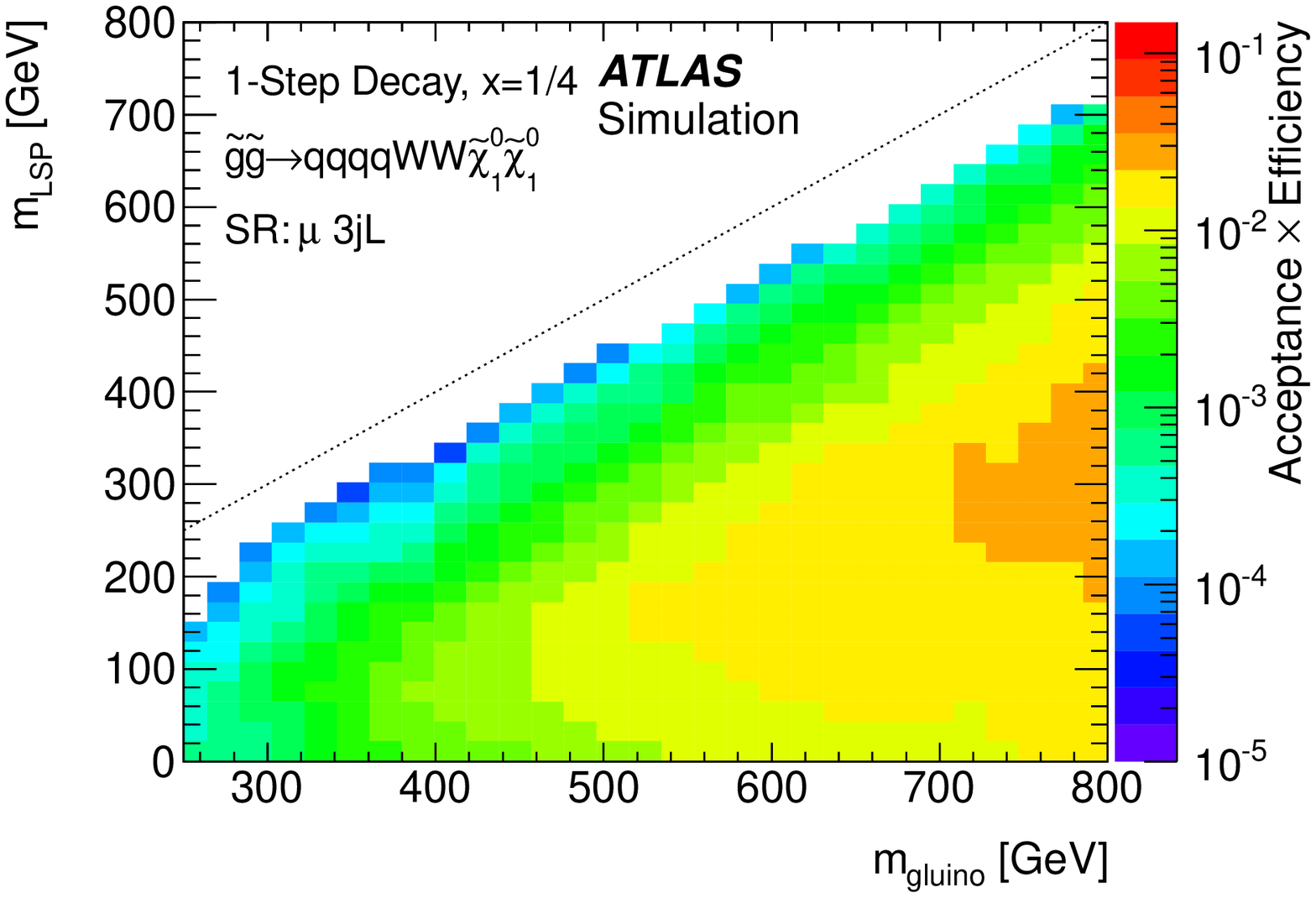}}
\quad\subfloat{
\includegraphics[width=0.45\textwidth]{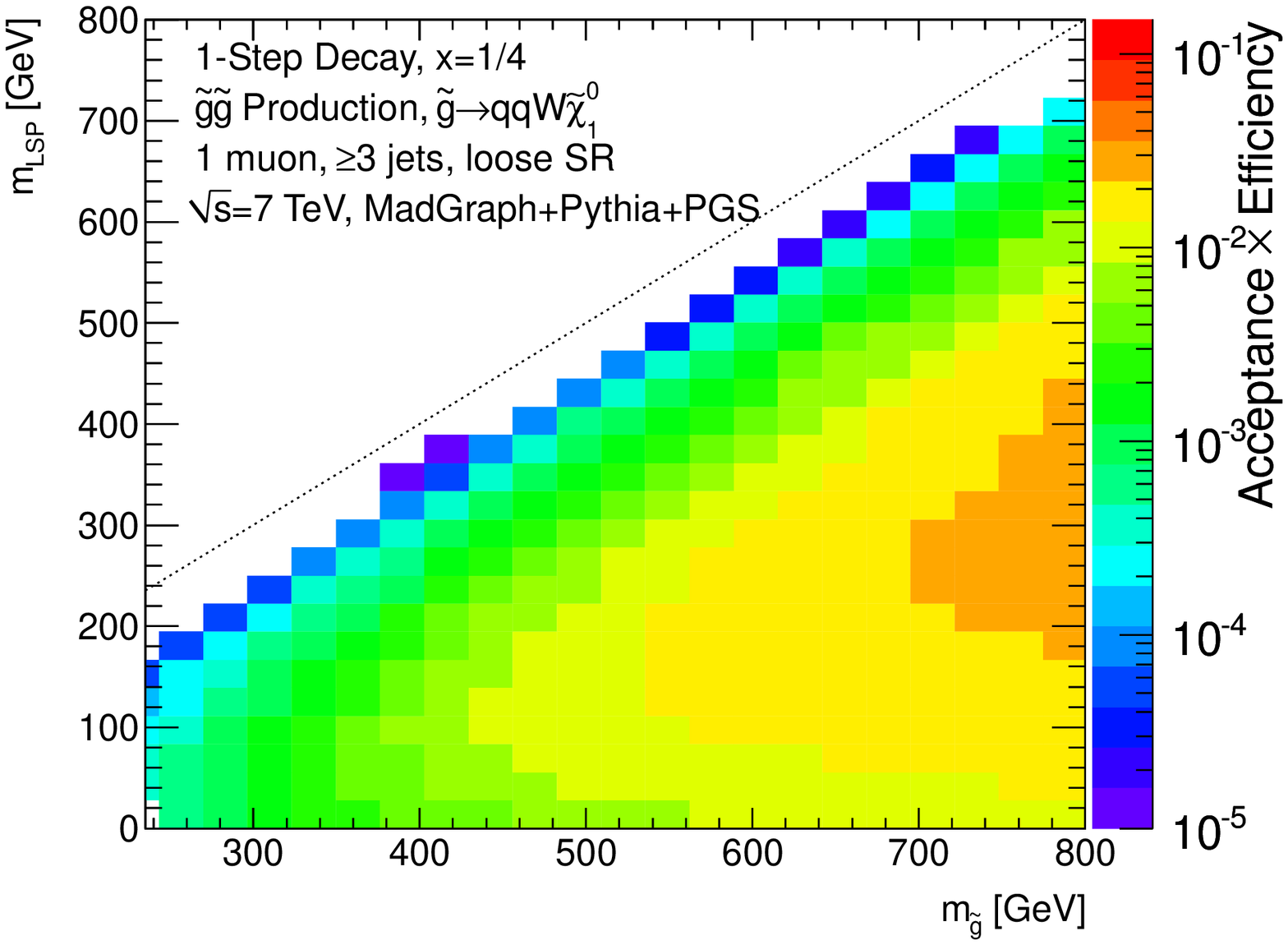}}
\caption{
Left, the public $A\epsilon$ for the ATLAS three jet ``loose'' one-lepton signal region from~\cite{1lpaper}.
Right, the same reproduced in the {\sc MadGraph} + {\sc Pythia} + {\sc PGS} setup used here.
Some differences are to be expected from the different generators and higher statistics used here, but the two follow one another closely.
}
\label{fig:smacceptancetimesefficiency}
\end{figure*}

In order to construct concrete limits with this method, the $A\epsilon$ for various simplified models must be made available by the LHC experiments.  While neither has provided tables of numbers, both CMS and ATLAS have published figures with the $A\epsilon$ for several models.  In order to demonstrate the value of publishing all such tables, we feel it is important to provide concrete limits that are comparable to those already published.  The $A\epsilon$ derived from {\sc PGS} is compared to that published by ATLAS in a simplified model grid in Figure~\ref{fig:smacceptancetimesefficiency}.  These results are sufficiently close to one another that, rather than wait for all results to be public, $A\epsilon$ results for the remaining grids are derived using {\sc PGS} and used directly in the remainder of this paper.  

Two conservative assumptions allow the inclusion of a larger number of production and decay modes.  The first is that for associated production the experimental $A\epsilon$ is at least as high as the $A\epsilon$ for the worse of the two production modes.  For inclusive searches this is generally a good assumption.  The minimum expected number of events would then be

\begin{eqnarray}
\label{eq:second}
N = \sigma_\text{tot} \times \mathcal{L}_\text{int} \times \sum_{(a,b)} P_{(a,b)} \times \\
\sum_i \text{BR}_{a\rightarrow i}\times \text{BR}_{b\rightarrow i} \times \min{ \left( A\epsilon_{a\rightarrow i} , A\epsilon_{b\rightarrow i} \right) }, \nonumber
\end{eqnarray}

\noindent where the first sum runs over all production modes, and only those where $a$ and $b$ are the same particle are included in Equation~\ref{eq:first}.  Similarly, the $A\epsilon$ for decays with different legs can be assumed to be at least as high as the $A\epsilon$ for the worse of the two legs.  That is,

\begin{eqnarray}
\label{eq:third}
N = \sigma_\text{tot} \times \mathcal{L}_\text{int} \times \sum_{(a,b)} P_{(a,b)} \times \sum_i \text{BR}_{a\rightarrow i}\times \\
\sum_j \text{BR}_{b\rightarrow j} \times \min{ \left( A\epsilon_{a\rightarrow i} , A\epsilon_{b\rightarrow j} \right) }, \nonumber
\end{eqnarray}

\noindent where diagrams with different decays on either side have now been included.

Two further assumptions would allow the setting of stricter limits.  One can assume that the experimental $A\epsilon$ for all production modes in the theory is similar to the average $A\epsilon$ for the production modes covered by simplified models.  In that case, the expected number of events can instead be written as

\begin{eqnarray}
\label{eq:fourth}
N = \sigma_\text{tot} \times \mathcal{L}_\text{int} \times \sum_{(a,a)} A\epsilon_{a\rightarrow i} \times \\
\frac{P_{(a,a)}}{\sum_{(a,a)} P_{(a,a)}} \times \text{BR}_{a\rightarrow i}^2 , \nonumber
\end{eqnarray}

\noindent where the sums are both over only those production modes covered by simplified models.  One might further assume that the $A\epsilon$ for all decay modes in the theory is similar to the average $A\epsilon$ for those events covered by the simplified model topologies.  Then the expected number of events may be written as:

\begin{eqnarray}
\label{eq:fifth}
N = \sigma_\text{tot} \times \mathcal{L}_\text{int} \times \\
\sum_i A\epsilon_{a\rightarrow i} \times \frac{P_{(a,a)}}{\sum_{(a,a)} P_{(a,a)}} \times \left(\frac{\text{BR}_{a\rightarrow i}} {\sum_{a} \text{BR}_{a\rightarrow i}}\right)^2 , \nonumber
\end{eqnarray}

\noindent where again the sums run only over the simplified models.  Clearly, the most aggressive mSUGRA limit is provided under this assumption, and a limit set in this manner risks claiming exlusion for regions which would not, in fact, be excluded at the 95\,\% confidence level by a dedicated search.  Although the accuracy of these two approximations might be suspect, the kinematic comparisons of simplified models to a complete SUSY parameter space point suggests that they are not completely unreasonable.

In order to portray the results achievable with present resources as accurately as possible, simplified model points were generated on a grid corresponding roughly to that already in use by the ATLAS experiment~\cite{1lpaper}.  Between these points, $A\epsilon$ is interpolated in the two-dimensional $m_\text{squark}$- / $m_\text{gluino}$-$m_\text{LSP}$ grid.  Because SM~3 and SM~4 are three dimensional grids, and because it is unlikely that experiments will provide full three-dimensional $A\epsilon$, three values of intermediate chargino mass are used: $m_\text{chargino} = x \times \left( m_\text{squark/gluino} - m_\text{LSP} \right) + m_\text{LSP}$, with $x=0.25$, $0.5$, and $0.75$.  To interpolate between these three two-dimensional planes, a simple quadratic fit is used.  When approaching the boundaries of $m_\text{LSP}=m_\text{chargino}$ and $m_\text{squark/gluino}=m_\text{chargino}$, the decay modes naturally turn off, making more complicated interpolation unnecessary.

\begin{figure*}
\centering
\graphicspath{{plots/}}
\subfloat[Combined zero-lepton SRs]{\label{fig:excl-official}
\includegraphics[width=0.47\textwidth]{modLimitLowTanBetaAccEff}}
\quad\subfloat[Simplified models]{\label{fig:excl-sm}
\includegraphics[width=0.47\textwidth]{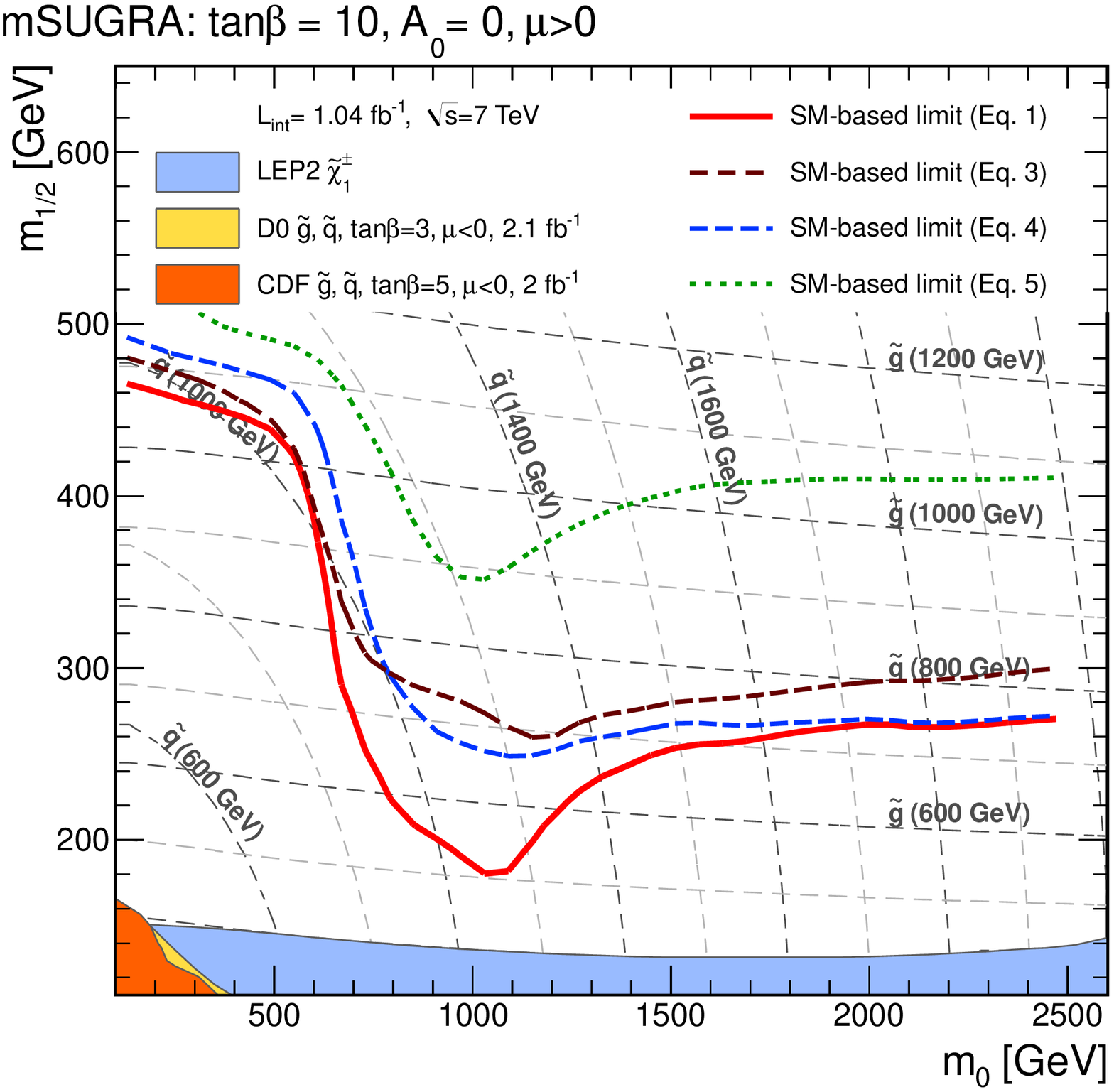}}
\caption{Combined zero-lepton exclusion limits for mSUGRA models with $\tan\beta=10$, $A_0=0$ and $\mu>0$ (\ref{fig:excl-official}) in comparison with the exclusion limit obtained using simplified models only (\ref{fig:excl-sm}). The signal region providing the best expected limit is taken for a given point in parameter space. The expected 95\,\% confidence level limit is shown as a dashed blue line, and the observed limit is shown as a solid red line. Results from previous searches are also shown for comparison purposes \cite{d-zero-1}--\cite{atlas-2010}, although some of these limits were produced using slightly different parameter choices. The simplified model limits are generated using four different sets of assumptions, corresponding to the limit equations in the main text.}
\label{fig:exclusion}
\end{figure*}

Using this procedure, limits are placed on mSUGRA using signal regions approximating the ATLAS zero-lepton search~\cite{0lpaper}. Five signal regions are included in this search, and the signal region with the best expected limit is used for each point. A point is considered to be excluded if the number of expected SUSY events in the optimal signal region exceeds the observed 95\,\% confidence level upper limit on new physics events in that signal region.  The results of the simplified model exclusion are compared to the zero-lepton exclusion without systematic uncertainties on the signal, as discussed previously, in Figure~\ref{fig:exclusion}.  Four simplified model exclusion curves are shown, corresponding to Equations~\ref{eq:first} and \ref{eq:third}--\ref{eq:fifth}.  In comparison to the zero-lepton exclusion limit, the most conservative simplified-model-based approach does rather poorly in the region dominated by $\tilde{q}\tilde{g}$ associated production, missing the correct limit by up to $\sim$100 GeV. This is in part due to the relatively complicated decay of the gluino (c.f.\ the large number of open modes in Figure~\ref{fig:br-plots2}). The coverage is much closer to the true limit for the region dominated by $\tilde{q}\tilde{q}$- and $\tilde{g}\tilde{g}$-production, for which the simplified model-derived limit is within 40 GeV of the true limit. 

From comparing the exclusion curves, one can indeed see that a conservative exclusion limit set using Eq.~\ref{eq:first} follows the ``correct'' exclusion limit quite well in regions of phase space that are well-covered by simplified models (c.f. Fig.~\ref{fig:smclassification}).  In regions that are not as well covered, Eq.~\ref{eq:third} still provides a conservative limit.  The aggressive limit set by Eq.~\ref{eq:fifth} overestimates the exclusion by up to 40 GeV in the squark-dominated region and by up to 100 GeV in the gluino-dominated region of phase space, because the assumption that the long gluino decay chains are well-modeled by the shorter chains of the simplified models is invalid at some level.  In terms of phase-space coverage, the conservative limits under-cover by 20\,\%, the middle two limits under-cover by 10\,\%, and the aggressive limit over-covers by 10\,\%.  Naturally, expanding the dictionary of simplified models available would improve the conservative limit and reduce the aggressive limit as more correct $A\epsilon$ are included for more production and decay modes. However, even with this small number of simplified models, the conservative limits set are close to the ``correct'' result.

\begin{figure*}
\centering
\graphicspath{{plots/}}
\quad\subfloat{\includegraphics[width=0.47\textwidth]{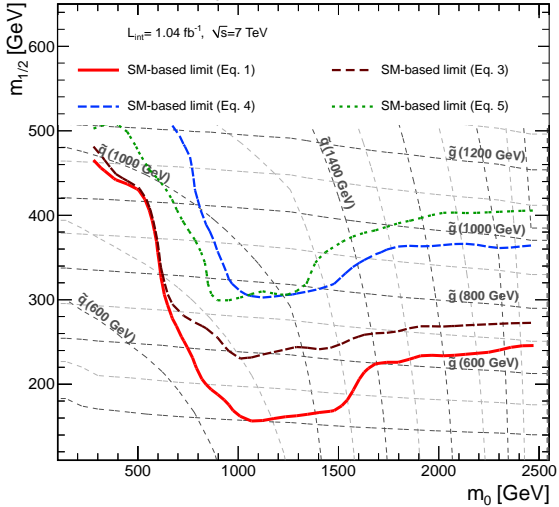}}
\quad\subfloat{\includegraphics[width=0.47\textwidth]{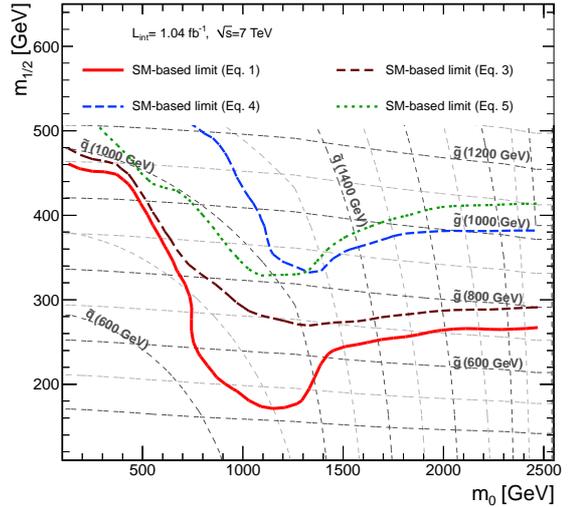}}
\caption{Exclusion limits for mSUGRA models with $\tan\beta=40$, $A_0=-500$ GeV and $\mu>0$ (left) and $\tan\beta=20$, $A_0=500$ GeV and $\mu>0$ (right) obtained using simplified models only. Combined limits are obtained by using the signal region which generates the best expected limit at each point in parameter space. The simplified model limits are generated using four different sets of assumptions, corresponding to the limit equations in the main text.}
\label{fig:exclusion2}
\end{figure*}

For demonstrative purposes, limits are also placed on an mSUGRA signal region at high $\tan(\beta)$.  The limits are shown in Figure~\ref{fig:exclusion2}.  Based on the agreement observed in Figure~\ref{fig:exclusion}, the experimental exclusion should lie a bit beyond the exclusion set by Eq.~\ref{eq:third}.

\subsubsection{Future Application}
\label{sec:future}

In extrapolating to more exotic theories, or even in expanding the applicability of a small list of simplified models to SUSY theories, several approximations can be made: that heavy-flavor jets are identical to light flavor jets for searches that do not include flavor tagging; that photons are identical to jets for searches that do not identify photons; or that more than half the time, chargino (neutralino) decays to the LSP via emission of a $W$-boson ($Z$-boson) produce a signature functionally identical to gluino decays via emission of two quarks.  Such approximations are physically well motivated and should result in limits which are in still better agreement with the full experimental results.

The method presented here would also benefit from understanding of the correlations between systematic uncertainties in all the available signal regions.  Here, five ATLAS signal regions were considered independently.  However, with correlation information one could combine all the ATLAS and CMS signal regions with public acceptance times efficiency data.  Even without correlations, one is free to choose the optimal signal region from either experiment for excluding a given point.  

\subsubsection{Summary and conclusion}
\label{sec:conclusion}
The application of simplified model limits to produce an exclusion contour in a complete SUSY model has been demonstrated.  Despite the apparent complexity of mSUGRA parameter space points, the kinematics can be well-reproduced by a combination of only a small number of simplified models.  The kinematic agreement is further improved when looking within a particular signal region, since the searches thus far conducted at the LHC tend to favor simplified model-like event topologies with a (relatively) small number of high-$p_T$ objects.  The exclusion contours derived from simplified models compare favorably with those already published with dedicated searches and, advantageously, dedicated Monte Carlo production.  The only additional information required for this recasting of results is the acceptance times efficiency for a sufficiently large dictionary of simplified models in the signal regions considered by the LHC experiments.  Once these data are available, it will be possible to trivially recast exclusions into more exotic SUSY theories, or even into non-SUSY theories with signatures covered by simplified models.  Much like the unfolding of Standard Model measurements, this method similarly allows a simple route for preservation of the data and application of current searches to future theories.

Practically, this approach means a significant resource saving for the LHC experiments.  By re-casting theories using information available from the matrix element and decay probabilities, no computing-intensive simulation of the model must be done.  Instead, the experiments are free to straightforwardly provide exclusion results in a large variety of theoretical models which include -- but may not be completely covered by -- simple final state signatures. Although the simplified models may not cover all the modes of a theory, with a relatively small number of simplified models it is possible to cover a fairly broad range of theories.  The exclusions acquired in this manner do not precisely overlap the results of a complete experimental search.  In the current LHC search era, however, they give a critical and surprisingly accurate estimation of how much theory space has already been excluded by the already conducted searches, and how much may still be open to discovery.  

\subsubsection*{Acknowledgements}
The authors would like to thank Jay Wacker for significant discussion of simplified models and potential pitfalls.  Many thanks also to Max Baak and Till Eifert for constructive criticism and encouragement whenever it was necessary. 
\renewcommand*{\refname}{\vspace*{-1.4cm}\subsubsection*{References}\vspace*{-4mm}}

\end{multicols}
%
%
%
\end{document}